\documentclass[11pt]{article}
\usepackage{amssymb}
\usepackage{graphics}
\usepackage{epsfig}
\usepackage{a4wide}
\usepackage{color}

\begin{document}

\newcommand{\ppww}{$pp \to W^+ W^-+X~$ }
\newcommand{\ppwwg}{$pp \to W^+ W^-g+X~$ }
\newcommand{\ppwwpol}{$pp \to W^+_{\lambda_3}W^-_{\lambda_4}+X~$}
\newcommand{\qqww}{$q\bar q \to W^+W^-~$}
\newcommand{\qqwwpol}{$q\bar q \to W^+_{\lambda_3}W^-_{\lambda_4}~$}
\newcommand{\qqwwg}{$q\bar q \to W^+W^-g~$}
\newcommand{\ggww}{$gg \to W^+W^-~$}
\newcommand{\ggwwpol}{$gg \to W^+_{\lambda_3}W^-_{\lambda_4}~$}
\newcommand{\ggwwg}{$gg \to W^+W^-g~$}
\newcommand{\qgwwq}{$q(\bar q)g \to W^+W^-q(\bar q)~$}
\newcommand{\ppuuww}{$pp (u\bar u) \to W^+W^-+X~$}
\newcommand{\ppqqww}{$pp (q\bar q) \to W^+W^-+X~$}
\newcommand{\ppggww}{$pp (gg) \to W^+W^-+X~$}
\newcommand{\ppuuwwg}{$pp (u\bar u) \to W^+W^-g+X~$}

\title{ $WW\gamma/Z$ production in the Randall-Sundrum model at the LHC and CLIC}
\author{ \textcolor{red}{}
Li Xiao-Zhou, Ma Wen-Gan, Zhang Ren-You, and Guo Lei\\
{\small Department of Modern Physics, University of Science and Technology}\\
{\small of China (USTC), Hefei, Anhui 230026, P.R.China}  }

\date{}
\maketitle \vskip 15mm
\begin{abstract}
We study the $W^+W^-\gamma(Z)$ productions at both the CERN Large
Hadron Collider (LHC) and the Compact Linear Collider (CLIC) in
the framework of the Randall-Sundrum (RS) model. The impacts of
the virtual RS Kaluza-Klein (KK) graviton on these processes are
studied and compared with the standard model (SM) background. We
present the integrated and differential cross sections in both the
RS model and the SM. The results show that the relative RS
discrepancies at the CLIC differ from those at the LHC,
particularly in the transverse momentum and rapidity
distributions. We also find that the RS signature performance, as
a result of the resonance character of the RS KK-graviton
spectrum, is distinctively unlike that in the large extra
dimensions model. We conclude that the CLIC with
unprecedented precision and high center-of-mass energy has a
potential advantage over the LHC in exploring the effects of the
RS KK graviton on the $W^+W^-\gamma(Z)$ production processes.
\end{abstract}


\vfill \eject

\baselineskip=0.32in

\renewcommand{\theequation}{\arabic{section}.\arabic{equation}}
\renewcommand{\thesection}{\Roman{section}.}
\newcommand{\nb}{\nonumber}

\newcommand{\Dir}{\kern -6.4pt\Big{/}}
\newcommand{\Dirin}{\kern -10.4pt\Big{/}\kern 4.4pt}
\newcommand{\DDir}{\kern -7.6pt\Big{/}}
\newcommand{\DGir}{\kern -6.0pt\Big{/}}

\makeatletter      
\@addtoreset{equation}{section}
\makeatother       

\vskip 5mm
\section{INTRODUCTION}
\par
Solving the huge disparity between the Planck scale $M_P$ and the
electroweak scale $M_{EW}$, which is known as the gauge hierarchy
problem, has long been the motivation for proposing new physics
beyond the standard model (SM). Strikingly distinct
from the supersymmetry or technicolor models, the extra dimensions
models, including the large extra dimensions (LED) \cite{1-LED}
model with factorizable geometry and the Randall-Sundrum (RS)
\cite{2-RS} model with nonfactorizable (warped) geometry, provide
alternative solutions to the gauge hierarchy problem by
postulating that the quantum gravity effects appear at the TeV
scale, which may induce rich collider phenomena at the CERN Large
Hadron Collider (LHC) and the future Compact Linear Collider
(CLIC) \cite{3-LHC-CLIC}.

\par
In the LED model \cite{1-LED}, we have $D=(4+n)$-dimensional
spacetime with $n$ being the number of extra dimensions
compactified on a $n$-dimensional torus with radius $R$, and the gauge
hierarchy problem is solved via the relation $M_P \sim R^{n/2}
M_D^{n/2 + 1}$ if $R$ is large enough. However, there appears a
new hierarchy between the $D$-dimensional fundamental scale $M_D\sim 1$~TeV
and the compactification radius $R^{-1}\sim $~eV-MeV in the LED
molel, which motivates proposing the RS model. The RS model is
based on a compactified warped extra dimension and two $3$-branes
in the background of the $\textrm{AdS}_5$ spacetime. In the RS model, the
gauge hierarchy problem is solved by an exponential warp factor.
The RS Kaluza-Klein (KK) graviton spectrum shares distinct
properties compared with that in the LED model, which has
inspired many works on the phenomenological studies in the RS
model, for example, the works on $p p \to V V,~\bar{l} l,~\bar{t}
t$ \cite{4-ppvv}-\cite{6-pptt}, $e^+ e^-, p p \to G_{KK} G_{KK}$
\cite{7-ppGkk}, and $e^+ e^-, p p \to \bar{l} l \gamma$
\cite{8-ppllr}.

\par
The triple gauge boson (TGB) productions are of particular
interest because they not only are sensitive to the quartic gauge
couplings (QGCs) but also could demonstrate new physics
signatures \cite{9-BSM}. Any deviation from the SM predictions
would hint at the existence of new physics, such as the non-SM
electroweak symmetry breaking mechanism or the extra
dimensions signals \cite{10-EWSB}. In this sense, the studies on
the TGB production channels in extra dimensions models, including
the LED model and the RS model, are necessary. Up to now, the TGB
productions have been thoroughly studied in the SM
\cite{11-ppvvvsm}, and the TGB production studies in the LED model
have also received impressive attention in the literature, including 
the neutral TGB production processes $pp \to
\gamma\gamma\gamma$, $pp \to \gamma\gamma Z$, $pp \to \gamma ZZ$
and $pp \to ZZZ$ in Ref.\cite{12-ppvvvled}, $e^+ e^- \to ZZZ$ at
the CLIC in Ref.\cite{13-eezzzled}, and $e^+ e^-, pp \to
W^+W^-\gamma/Z$ at the LHC and ILC in Ref.\cite{14-wwvled}. In the RS model,
only the TGB production process $pp \to
\gamma\gamma\gamma$ has been studied in Ref.\cite{15-rrrRS}.

\par
In the present paper, we consider the effects of the virtual RS
KK-graviton exchange on the $W^+W^-\gamma$ and $W^+W^-Z$
productions at both the LHC and the CLIC. Three areas of interest
motivate this work. First, the $W^+W^-\gamma$ and
$W^+W^-Z$ productions are excellent probes of the SM QGCs.
Second, and different from the LED model, the fact that the RS
KK-graviton spectrum generally manifests itself as TeV-order
resonances could alter the cross sections and thus lead to
identifiable changes in the TGB phenomenology at the LHC and the
multi-TeV CLIC. Third, compared with the current data for QGCs
available from LEP II and Tevatron \cite{16-LEP-Tevatron}, the LHC
can provide more precise measurements of the QGCs due to its
high energy and luminosity \cite{17-LHC}, and the multi-TeV CLIC
can probe the QGCs with unprecedented precision due to the cleaner
environment arising from $e^+e^-$ collisions and the compelling
high energy \cite{18-CLIC}. Therefore, it can be expected that the
LHC and the future multi-TeV CLIC will provide complementary
studies on the TGB productions. The rest of the paper is organized as
follows. In Sec.II, we briefly describe the related theory of
the RS model. In Sec.III, the calculation strategy is
presented. We perform the numerical analyses for the
$W^+W^-\gamma$ and $W^+W^-Z$ productions at both colliders in
Sec.IV. Section V is devoted to a short summary. In the Appendix,
we present the relevant Feynman rules for vertices of the RS
KK graviton coupled with the SM fields.

\vskip 5mm
\section{ RELATED THEORY }
\label{related theories}
\par
In the brane-bulk scenario of the RS model \cite{2-RS}, the
spacetime is assumed to be $5=(4+1)$-dimensional with the one-dimensional
extra dimension compactified on an $S_1/Z_2$ orbifold with radius $R_c$,
and two $3$-branes, the Planck brane and TeV brane, reside at the
orbifold fixed points $\phi=0,~\pi$, respectively. The
five-dimensional bulk connecting the two branes is a slice of the
$\textrm{AdS}_5$ spacetime, which is nonfactorizable and has a constant
negative curvature. The nonfactorizable bulk metric is given by
\begin{equation}
ds^2 = e^{-{\cal K}R_c\phi}\eta_{\mu\nu}dx^{\mu}dx^{\nu}~+~R_c^2d\phi^2 ,
\end{equation}
where $0\le \phi \le \pi$, and ${\cal K}$ is the curvature scale of the
bulk. It is assumed that the SM fields are located at
the TeV brane, while the gravity can propagate in the whole five-dimensional
bulk. The Planck scale $M_P$ for gravity can be suppressed to
the TeV scale via the exponential warp factor $e^{-\pi {\cal K} R_c}$,
i.e., $M_P e^{-\pi {\cal K} R_c}\sim {\cal O}({\rm TeV})$ and thus the
gauge hierarchy problem is solved. In the low-energy effective
four-dimensional theory view, after taking a linear expansion of the
gravity field as fluctuations around the flat metric and adopting
the KK reduction \cite{19-RSL}, we obtain the interactions between
the KK tower of massive spin-2 gravitons and the SM particles as
\begin{equation} \label{Lagrangian}
{\cal L}  =  -{1 \over \overline{M}_P} T^{\mu\nu}(x)
G_{KK,\mu\nu}^{(0)}(x) -{1\over\Lambda_\pi} T^{\mu\nu}(x)\sum_{n=1}^{\infty}
G_{KK,\mu\nu}^{(n)}(x) ,
\end{equation}
where $\overline{M}_P=M_P/\sqrt{8 \pi}$ is the reduced Planck scale and
$T^{\mu\nu}$ is the energy-momentum tensor of the SM particles.
The interactions between the zero mode of the RS KK graviton
and the SM particles are suppressed by $\overline{M}_P$ and thus
decouple, while the couplings of the
massive RS KK graviton are suppressed by $\Lambda_\pi = e^{-\pi {\cal K} R_c}
\overline{M}_P \sim$TeV. The mass of the $n$th RS KK graviton
$G_{KK,\mu\nu}^{(n)}$ is
\begin{equation}
M_n = x_n {\cal K} ~e^{-\pi {\cal K} R_c} = {x_n \over x_1} M_1 ,
\end{equation}
where $x_n$ are roots of the equation of Bessel function $J_1(x)$,
i.e., $J_1(x_n)=0$. For example, the first three roots are $x_1
\simeq 3.83$, $x_2 \simeq 7.02$, and $x_3 \simeq 10.17$. That
shows the masses of the RS KK gravitons are unevenly spaced with
the mass splitting of TeV order if $M_1 \sim {\rm TeV}$.

\par
The two independent input parameters in the RS model are chosen as
\begin{eqnarray}
M_1 = x_1 {\cal K} e^{-\pi {\cal K} R_c} , ~~~~~
c_0 = {\cal K}/\overline{M}_P ,
\end{eqnarray}
where $M_1$ is the mass of the first RS KK graviton and $c_0$ is the
effective coupling. The theoretical requirements \cite{19-RSL,
20-c0} constrain $c_0$ in the range of $0.01\le c_0 \le 0.1$.
The Feynman rules for the RS model can be read off from the
counterparts in the LED model \cite{21-LEDrules} upon the
replacement of \cite{8-ppllr, 19-RSL, 22-RSbible}
\begin{equation}
{\kappa \over 2} \to {1 \over \Lambda_\pi} = {c_0 \over M_1} x_1 ,
\end{equation}
where $\kappa = {2 \over \overline{M}_P}$ is the gravitational
coupling strength in the LED model \cite{21-LEDrules}. We present
the explicit expressions of the vertex Feynman rules in the RS
model related to our calculations in the Appendix. The spin-2 RS
KK-graviton propagator in the de Donder gauge can be expressed as
\begin{eqnarray}
\tilde{G}_{\rm KK}^{\mu \nu \alpha \beta}=\frac{1}{2} D(\hat{s})
\left[\eta^{\mu \alpha} \eta^{\nu \beta} + \eta^{\mu \beta}
\eta^{\nu \alpha} - \frac{2}{1+2}\eta^{\mu \nu} \eta^{\alpha \beta}
\right] ,
\end{eqnarray}
where the summation over the tower of the KK graviton is
\begin{equation} \label{D(s)}
D(\hat{s}) = \sum_{n=1}^\infty \frac{i}{\hat{s}- M_{n}^2+iM_n \Gamma_n}  .
\end{equation}
The total decay width of the $n$th KK graviton can be expressed as \cite{6-pptt,22-RSbible}
\begin{equation}
 \Gamma_n = {1 \over 8\pi}x_n^2M_nc_0^2\Delta_n ,
\end{equation}
and
\begin{equation}
\Delta_n = \Delta_n^{\gamma \gamma} + \Delta_n^{gg}
         + \Delta_n^{WW} + \Delta_n^{ZZ}
         + \sum_\nu \Delta_n^{\nu\nu} + \sum_l \Delta_n^{ll}
         + \sum_q \Delta_n^{qq}
         + \Delta_n^{HH} ,
\end{equation}
where $\Delta_n^{yy}$ is a numerical coefficient for the decay
$G^{(n)}_{KK}(x) \to yy$, and $y$ is the SM particle involved. The
explicit values for $\Delta_n^{yy}$ are given in Refs.\cite{6-pptt,
8-ppllr, 21-LEDrules}.

\par
From Eqs.(\ref{Lagrangian}) and (\ref{D(s)}), one can find that all the
massive RS KK gravitons should be
considered and summed over. Howerver, the contributions of the higher
modes of the massive KK gravitons are negligible due to the fact that the
higher zeros of the Bessel function $J_1(x)$ generate heavier
RS KK gravitons with masses of several TeV \cite{22-RSbible}.
For simplicity, we consider only the lightest RS KK graviton ($n = 1$)
resonance which provides the dominant contribution \cite{8-ppllr, 23-lightest}.

\vskip 5mm
\section{ CALCULATION STRATEGY }
\label{calculations}
\par
\begin{figure*}
\begin{center}
\includegraphics[scale=0.8]{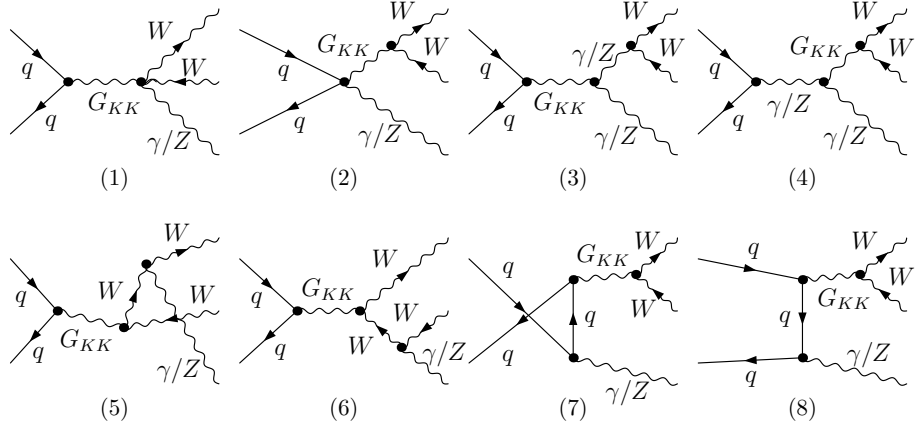}
\caption{\label{fig1} The LO Feynman diagrams for the partonic
process $q\bar q \to W^+W^-\gamma/Z~(q = u, d, s, c, b)$ with
KK-graviton exchange in the RS model. The SM-like diagrams are not
shown.}
\end{center}
\end{figure*}
\begin{figure*}
\begin{center}
\includegraphics[scale=0.8]{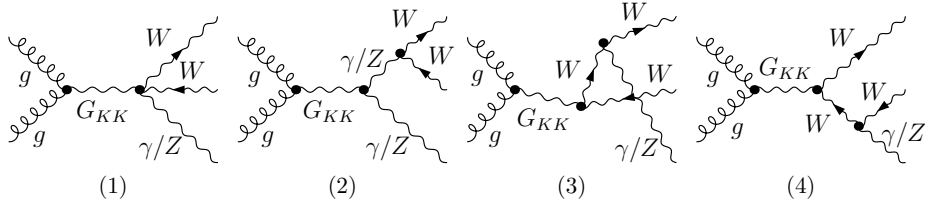}
\caption{\label{fig2} The LO Feynman diagrams for the gluon-gluon
fusion subprocess $g g \to W^+W^-\gamma/Z$ with KK-graviton exchange
in the RS model. }
\end{center}
\end{figure*}
\begin{figure*}
\begin{center}
\includegraphics[scale=0.8]{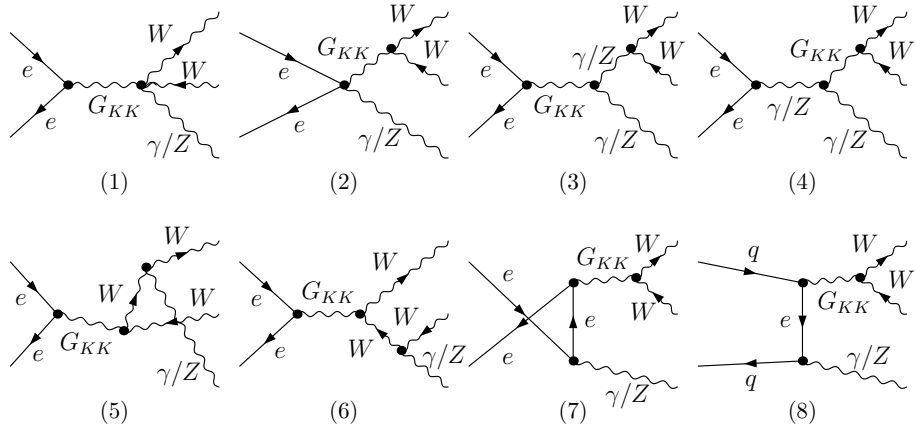}
\caption{\label{fig3} The LO Feynman diagrams for the $e^+e^- \to W^+W^-\gamma/Z$
process with KK-graviton exchange in the RS model. The SM-like diagrams are not shown.}
\end{center}
\end{figure*}

The calculation strategy in this section is similar with that in
Ref.\cite{14-wwvled}. The $pp \to W^+W^-\gamma, W^+W^-Z$ processes
at the LHC include two kinds of subprocesses: the quark-antiquark
annihilation and the gluon-gluon fusion, which are denoted as
\begin{eqnarray}\label{channel-1}
~~~q(p_1)+\bar q(p_2) \to W^{+}(p_3)+W^{-}(p_4)+V(p_5),~~~~~~~~~~~~~~~~~\\
\label{channel-2} g(p_1)+g(p_2) \to
W^{+}(p_3)+W^{-}(p_4)+V(p_5),~~~~~~~~~~~~~~~~~
\end{eqnarray}
where $q = u, d, s, c, b$ and $V=\gamma,Z$. The $e^+ e^- \to
W^+W^-\gamma, W^+W^-Z$ processes at the CLIC can be denoted as
\begin{eqnarray}\label{channel-3}
e^+(p_1)+e^-(p_2) \to  W^{+}(p_3)+W^{-}(p_4)+V(p_5),
~~~~(V=\gamma,Z).
\end{eqnarray}
In reactions (\ref{channel-1}), (\ref{channel-2}) and
(\ref{channel-3}), $p_{i}$ $(i=1,2,3,4,5)$ are the four-momenta of
the incoming and outgoing particles. The leading order (LO) Feynman
diagrams with RS KK-graviton exchange for the partonic processes
(\ref{channel-1}) and (\ref{channel-2}) are depicted in
Figs.\ref{fig1} and \ref{fig2}, respectively, while the LO non-SM-like
Feynman diagrams in the RS model for the process (\ref{channel-3}) are
depicted in Fig.\ref{fig3}.

\par
From the Feynman diagrams shown in Figs.\ref{fig1}-\ref{fig3}, one
can find that the RS KK graviton couples not only to the fermion
pair ($f \bar{f} G_{KK}$), vector boson pair ($V V G_{KK}$), and
fermion-antifermion-vector boson ($f\bar f V G_{KK}$) but also to
the TGB vertices ($V V V G_{KK}$), which is similar to the case
in the LED model \cite{14-wwvled}. In this sense, it is natural to
expect that the RS KK graviton with mass of TeV order may induce
considerably distinctive effects on the TGB production processes
at the LHC and the future multi-TeV CLIC.

\par
We express the Feynman amplitudes for the subprocesses $q
\bar{q} \to W^+W^-\gamma/Z$ $(q=u,d,c,s,b)$ and $gg \to W^+W^-\gamma/Z$ as
\begin{equation}
{\cal M}_{q\bar{q}}^{\gamma/Z} = {\cal M}_{q\bar{q}}^{\gamma/Z,SM} +
{\cal M}_{q\bar{q}}^{\gamma/Z,RS},~~~~{\cal M}_{gg}^{\gamma/Z} =
{\cal M}_{gg}^{\gamma/Z,RS},
\end{equation}
where ${\cal M}_{q\bar{q}}^{\gamma/Z,SM}$ stands for
the amplitude contributed by the SM-like diagrams, while ${\cal
M}_{q\bar{q}}^{\gamma/Z,RS}$ and ${\cal M}_{gg}^{\gamma/Z,RS}$ are
the amplitudes contributed by the diagrams with virtual RS KK-graviton
exchange. The Feynman amplitude for the $e^+ e^- \to W^+W^-\gamma/Z$
process can be separated into two components,
\begin{equation}
{\cal M}_{ee}^{\gamma/Z} = {\cal M}_{ee}^{\gamma/Z,SM} + {\cal
M}_{ee}^{\gamma/Z,RS},
\end{equation}
where ${\cal M}_{ee}^{\gamma/Z,SM}$ and ${\cal
M}_{ee}^{\gamma/Z,RS}$ represent the amplitudes of the SM-like and the RS KK-graviton exchange diagrams, respectively.

\par
The total cross sections for the partonic processes $q \bar{q}(gg) \to
W^+W^-\gamma/Z$ have the form 
\begin{eqnarray}\label{int-ppvvv}
\hat{\sigma}_{ij}^{\gamma/Z} =
\frac{1}{4|\vec{p}|\sqrt{\hat{s}}}\int {\rm d}\Gamma_3
\sum_{spin}^{color}{}^\prime|{\cal
M}_{ij}^{\gamma/Z}|^2,
~~~~(ij=u\bar{u},d\bar{d},c\bar{c},s\bar{s},b\bar{b},gg),
\end{eqnarray}
where $\vec{p}$ is the three-momentum of one initial parton in the
center-of-mass system (c.m.s), the summation is taken over the spins
and colors of the initial and final states, and the prime on the sum
denotes averaging over the initial spins and colors. The three-body
phase space element ${\rm d}\Gamma_3$ is defined as
\begin{eqnarray}
{\rm d} \Gamma_3 = (2\pi)^4  \delta^{(4)} \left(
p_1+p_2-\sum_{i=3}^5 p_i \right) \prod_{i=3}^5 \frac{d^3
\vec{p}_i}{(2\pi)^3 2E_i}.
\end{eqnarray}

\par
The total cross sections for the $pp \to W^+W^-\gamma/Z$ processes at
the hadronic level are obtained by convoluting
$\hat{\sigma}^{\gamma/Z}_{i j}$ with the parton distribution
functions (PDFs) of the colliding protons in the following way,
\begin{eqnarray}
\sigma_{pp}^{\gamma/Z} = \sum^{c\bar c,b\bar b,gg}_{ij=u\bar
u,d\bar d,s\bar s}\frac{1}{1+\delta_{ij}} \int dx_A dx_B \left[
G_{i/A}(x_A,\mu_f)
G_{j/B}(x_B,\mu_f)\hat{\sigma}^{\gamma/Z}_{ij}(\sqrt{\hat{s}}=x_Ax_B\sqrt{s})
+(A \leftrightarrow B) \right], \nb \\
\end{eqnarray}
where $G_{i/P}$ $(i = q, \bar{q}, g, P=A,B)$ represents the PDF of
parton $i$ in proton $P(=A,B)$, $\mu_{f}$ is the factorization
scale, and $x_A$ and $x_B$ refer to the momentum fractions of the parton
(quark or gluon) in protons $A$ and $B$, respectively. The total
cross sections for $e^+ e^- \to W^+W^-\gamma(Z)$ can be expressed
as
\begin{eqnarray}\label{int-eevvv}
\sigma_{ee}^{\gamma/Z} &=& \frac{1}{4|\vec{p}|\sqrt{s}}\int {\rm
d}\Gamma_3 \sum_{spin}{}^\prime|{\cal M}_{ee}^{\gamma/Z}|^2,
\end{eqnarray}
where $\vec{p}$ is the three-momentum of the incoming $e^+$ (or $e^-$) in
the c.m.s of the $e^+e^-$ collider. The prime on the
sum means averaging over the initial spin states as declared for Eq.(\ref{int-ppvvv}).

\vskip 5mm
\section{NUMERICAL RESULTS AND DISCUSSIONS}
\par
In this section we present the numerical results and the kinematic
distributions for the $W^+W^-\gamma$ and $W^+W^-Z$ productions in
both the SM and the RS model at the LHC and the CLIC. For the
computations at the LHC, we use the CTEQ6L1 PDFs \cite{24-PDF} with
$\Lambda_{QCD} = 165$~MeV and $n_{f}=5$ and take the factorization
scale as $\mu_f=m_W$ and $\mu_f=m_W + m_Z/2$ for the $pp \to W^+ W^-
\gamma$ and $pp \to W^+ W^- Z$ processes, respectively. The masses
of the active quarks are neglected, i.e., $m_{q}=0~(q=u,d,c,s,b)$, and
the CKM matrix is set to be the unit matrix. The other relevant
input parameters are chosen as \cite{25-PDG}
\begin{eqnarray}
~~~~~~\alpha^{-1}(0)=137.036,~~m_W=80.385~\textrm{GeV},~~m_Z=91.1876~\textrm{GeV}, \nb \\
M_{H}=125~\textrm{GeV},~~m_{t}=173.5~\textrm{GeV},~~m_{e}=0.511 \times 10^{-3}~\textrm{GeV}.
\end{eqnarray}
For the $e^+e^-/pp \to W^+W^-\gamma$
processes we apply a transverse momentum cut
$p_{T}^{\gamma}>25$~GeV and a rapidity cut $|\eta^{\gamma}|<2.7$ on
the final photon in order to remove the infrared (IR) singularity at the tree
level.

\par
Recently, the dilepton searches at ATLAS have excluded at the $95\%$
confidence level the RS KK graviton with masses below $2.16$~TeV
\cite{26-dilepton}. The diphoton experiments at ATLAS provided
$95\%$ confidence level lower limits on the lightest RS KK-graviton
mass $M_1$ \cite{27-diphoton}: $1.03$~TeV for $c_0=0.01$, and
$2.23$~TeV for $c_0=0.1$. In the present paper, we choose
$M_1=2.25$~TeV and $c_0=0.1$ unless otherwise stated.

\par
\begin{figure}[htbp]
\begin{center}
\includegraphics[scale=0.7]{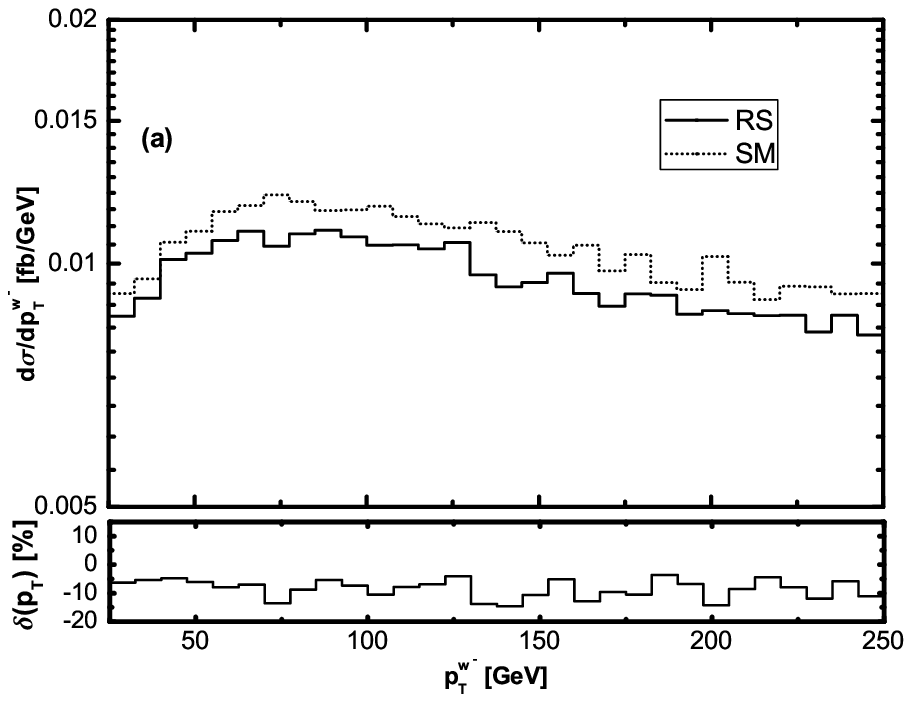}%
\hspace{0in}%
\includegraphics[scale=0.7]{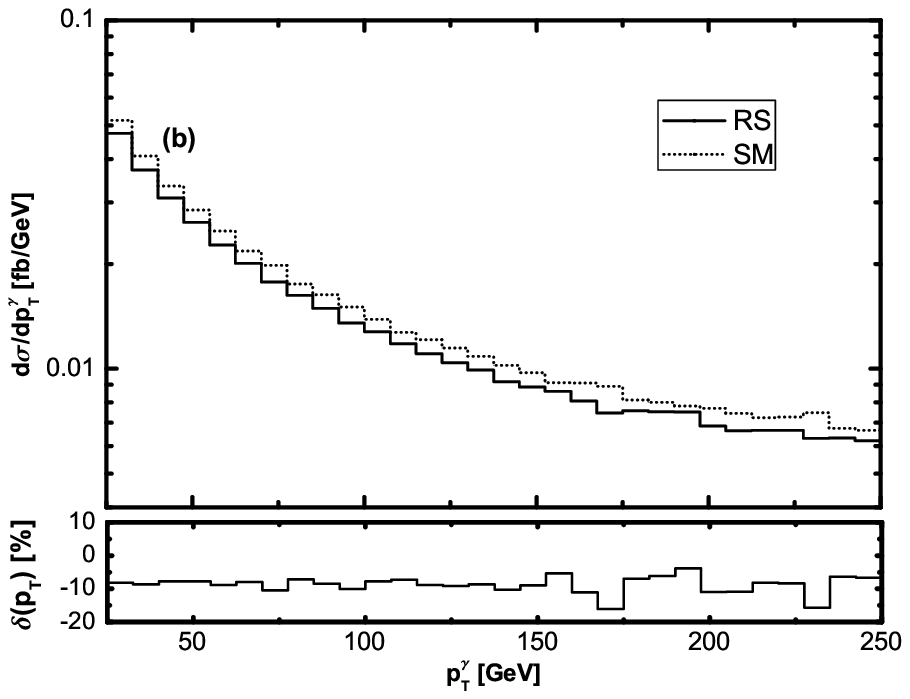}%
\hspace{0in}%
\caption{ \label{fig4} The transverse momentum distributions of the final $W^-$ and
$\gamma$ and the corresponding relative RS discrepancies for the $e^+e^- \to W^{+}W^{-}\gamma$
process at the $\sqrt{s}=5$~TeV CLIC, with the RS parameters $M_1=2.25$~TeV and
$c_0=0.1$. (a) for $p_T^{W^-}$, (b) for $p_T^{\gamma}$. }
\end{center}
\end{figure}
\begin{figure}[htbp]
\begin{center}
\includegraphics[scale=0.7]{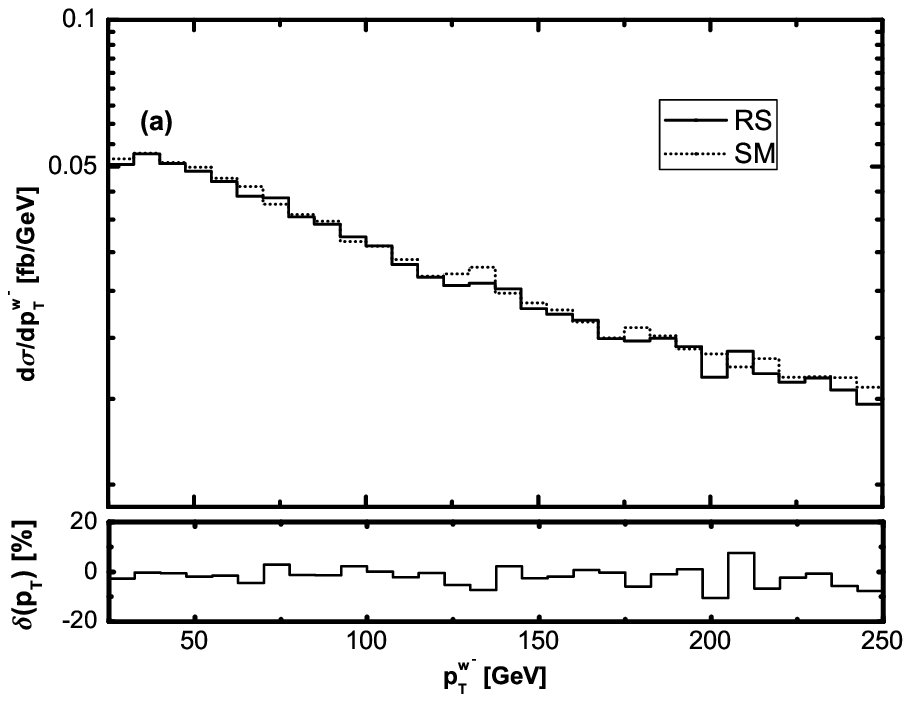}%
\hspace{0in}%
\includegraphics[scale=0.7]{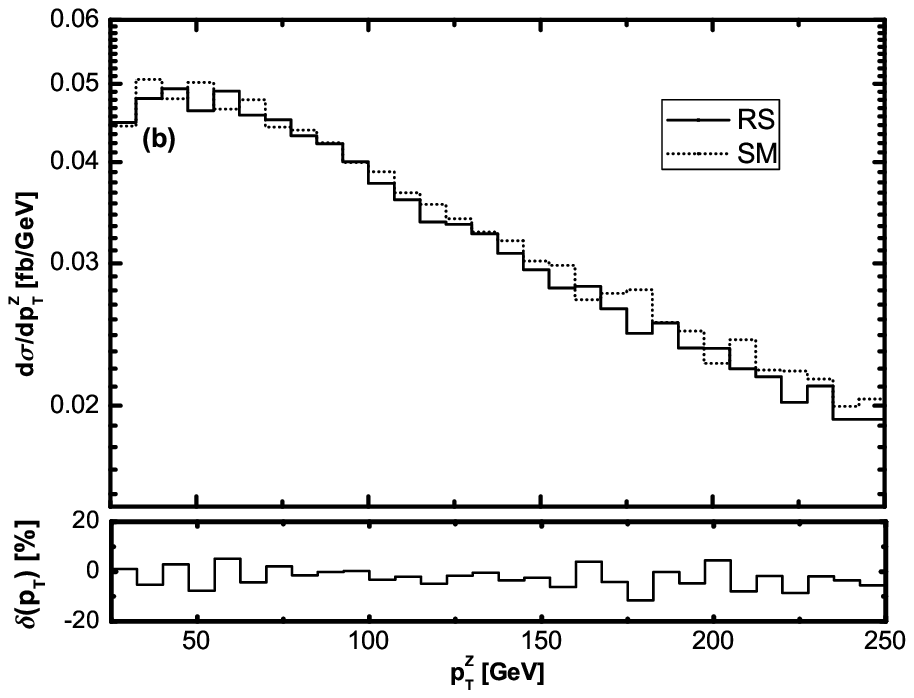}%
\hspace{0in}%
\caption{ \label{fig5}  The transverse momentum distributions of the final $W^-$ and
$Z$ and the corresponding relative RS discrepancies for the $e^+e^- \to W^{+}W^{-}Z$
process at the $\sqrt{s}=5$~TeV CLIC, with the RS parameters $M_1=2.25$~TeV and
$c_0=0.1$. (a) for $p_T^{W^-}$, (b) for $p_T^{Z}$. }
\end{center}
\end{figure}
\begin{figure}[htbp]
\begin{center}
\includegraphics[scale=0.7]{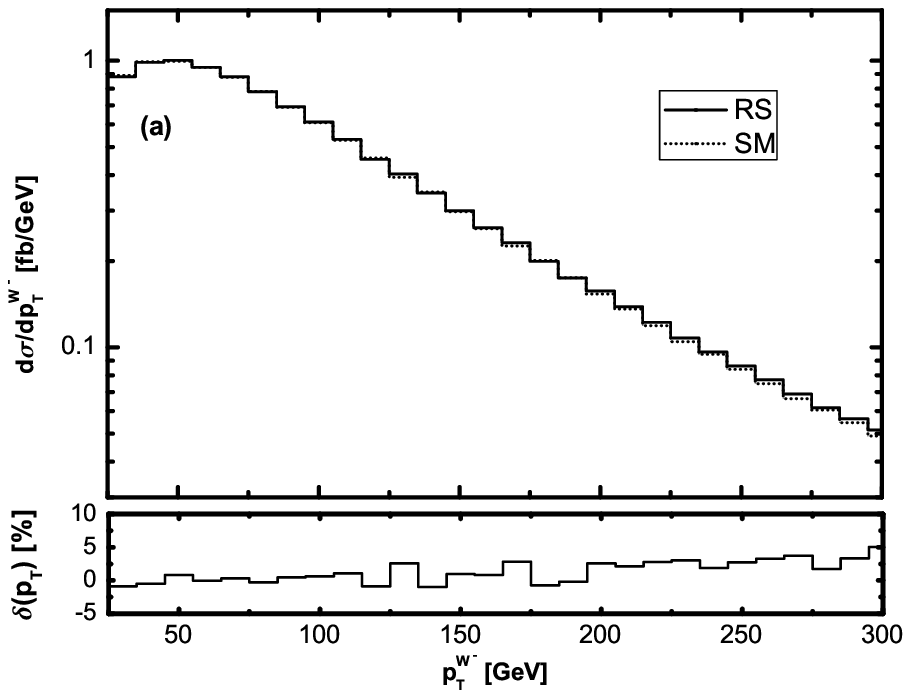}%
\hspace{0in}%
\includegraphics[scale=0.7]{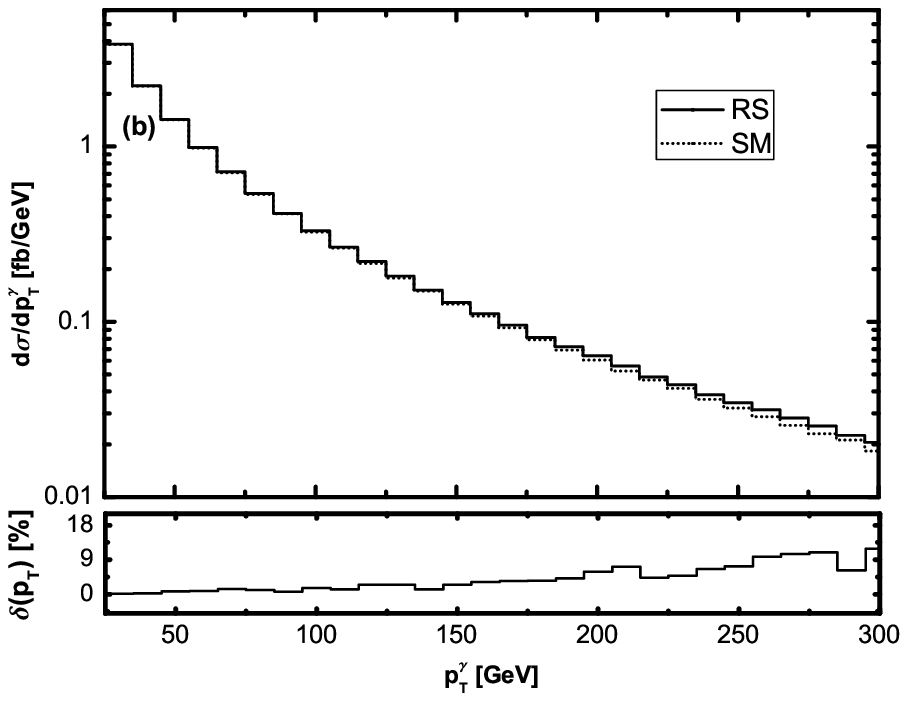}%
\hspace{0in}%
\caption{ \label{fig6} The transverse momentum distributions of the final $W^-$ and
$\gamma$ and the corresponding relative RS discrepancies for the $pp \to W^{+}W^{-}\gamma$
 process at the $\sqrt{s}=14$~TeV LHC, with the RS parameters $M_1=2.25$~TeV and
$c_0=0.1$. (a) for $p_T^{W^-}$, (b) for $p_T^{\gamma}$. }
\end{center}
\end{figure}
\begin{figure}[htbp]
\begin{center}
\includegraphics[scale=0.7]{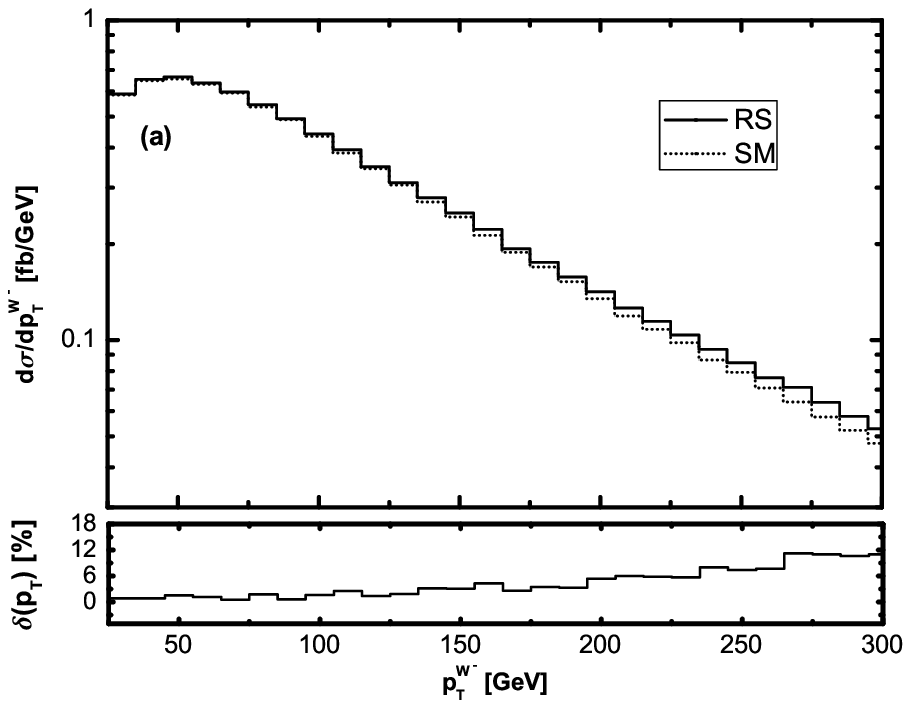}%
\hspace{0in}%
\includegraphics[scale=0.7]{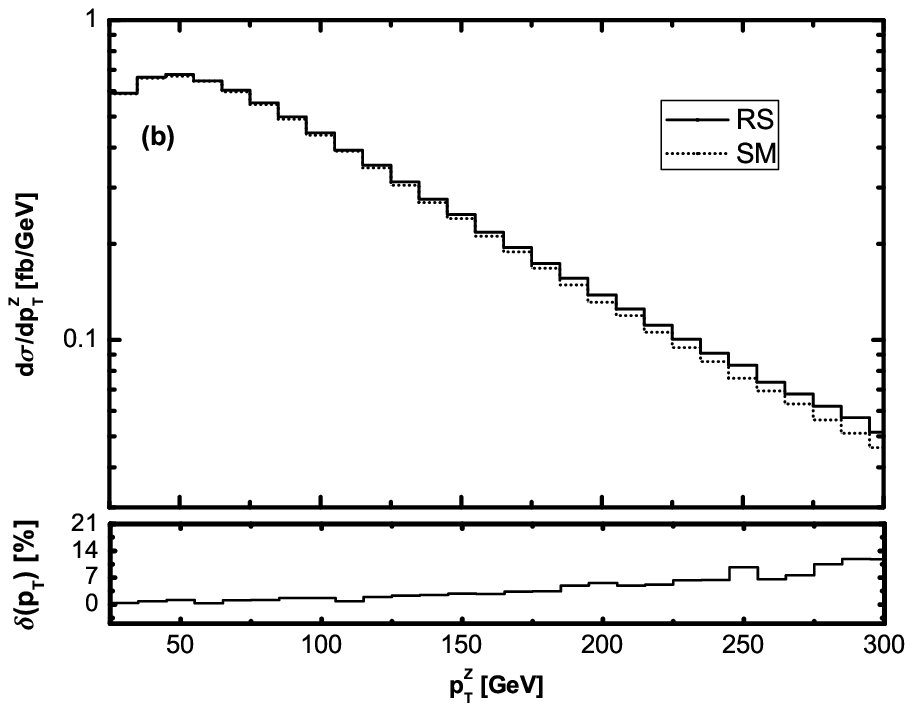}%
\hspace{0in}%
\caption{ \label{fig7} The transverse momentum distributions of the final $W^-$ and
$Z$ and the corresponding relative RS discrepancies for the $pp \to W^{+}W^{-}Z$
process at the $\sqrt{s}=14$~TeV LHC, with the RS parameters $M_1=2.25$~TeV and $c_0=0.1$. (a) for
$p_T^{W^-}$, (b) for $p_T^{Z}$. }
\end{center}
\end{figure}

In Figs.\ref{fig4} and \ref{fig5}, we depict the transverse momentum
($p_T$) distributions of final $W^{-}$, $\gamma$ and $Z$ for the
$e^+e^- \to W^+W^-\gamma, W^+W^-Z$ processes at the
$\sqrt{s}=5$~TeV CLIC, respectively. Figures \ref{fig6} and
\ref{fig7} show the $p_T^{W^-}$, $p_T^{\gamma}$ and $p_T^Z$
distributions for the processes $pp \to W^+W^-\gamma, W^+W^-Z$ at
the $\sqrt{s}=14$~TeV LHC. In each plot of
Figs.\ref{fig4}-\ref{fig7}, we provide $p_T^{W^-}$, $p_T^{\gamma}$
and $p_T^Z$ distributions in both the SM and the RS model for
comparison. We define the relative RS discrepancy of $p_T$
distribution as $\delta(p_T)\equiv \left(\frac{d\sigma_{RS}}{dp_T}-
\frac{d\sigma_{SM}}{dp_T}\right)/\frac{d\sigma_{SM}}{dp_T}$ to
describe the virtual RS KK-graviton effect, and plot the corresponding
$\delta(p_T)$ distribution in the nether plot for each figure. From
Fig.\ref{fig4} we find that at the CLIC both $\delta(p_T^{W^-})$ and
$\delta(p_T^{\gamma})$ for the $e^+e^- \to W^+W^-\gamma$ process
lie in the negative range, which means that the RS KK-graviton
mediated processes attenuate the SM background in the $p_T$ region
of $25$~GeV $\le p_T \le 250$~GeV. The curves of
$\delta(p_T^{W^-})$ and $\delta(p_T^Z)$ for the $e^+e^- \to W^+W^-Z$
process behave in a similar way as the former process. Figures \ref{fig6}
and \ref{fig7} show that the RS effect at the LHC enhances the SM
contributions in the same region, and the relative discrepancies
$\delta(p_T^{W^-})$, $\delta(p_T^{\gamma})$ and $\delta(p_T^{Z})$ at
the LHC become larger with the increment of $p_T$. The
distinct characteristic of the $p_T$ distributions at the CLIC and
the LHC can serve as the complementary study on the TGB production
events.

\par
\begin{figure}[htbp]
\begin{center}
\includegraphics[scale=0.7]{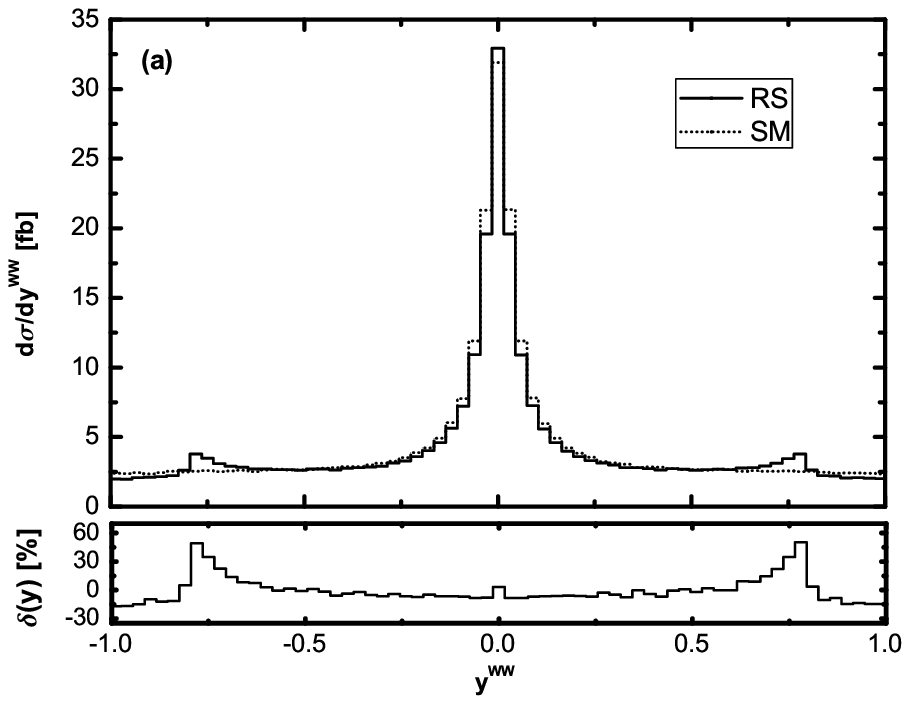}%
\hspace{0in}%
\includegraphics[scale=0.7]{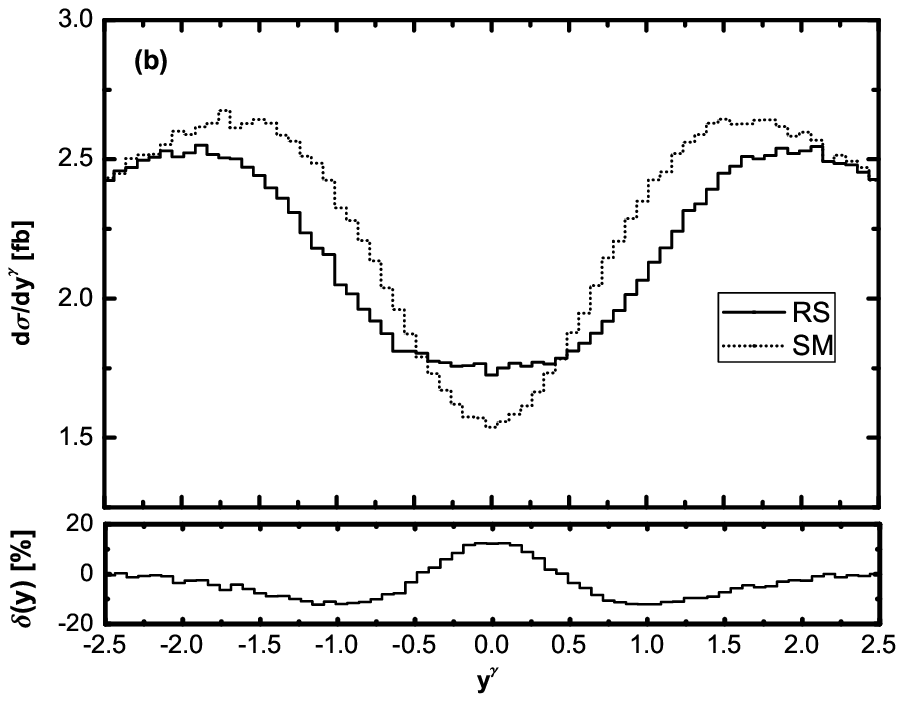}%
\hspace{0in}%
\caption{ \label{fig8} The rapidity distributions of the final $W$ pair and
$\gamma$ and the corresponding relative RS discrepancies for the
$e^+e^- \to W^{+}W^{-}\gamma$ process in both the SM and the RS
model at the $\sqrt{s}=5$~TeV CLIC, with the RS parameters
$M_1=2.25$~TeV and $c_0=0.1$, (a) for $y^{WW}$. (b) for
$y^{\gamma}$. }
\end{center}
\end{figure}
\begin{figure}[htbp]
\begin{center}
\includegraphics[scale=0.7]{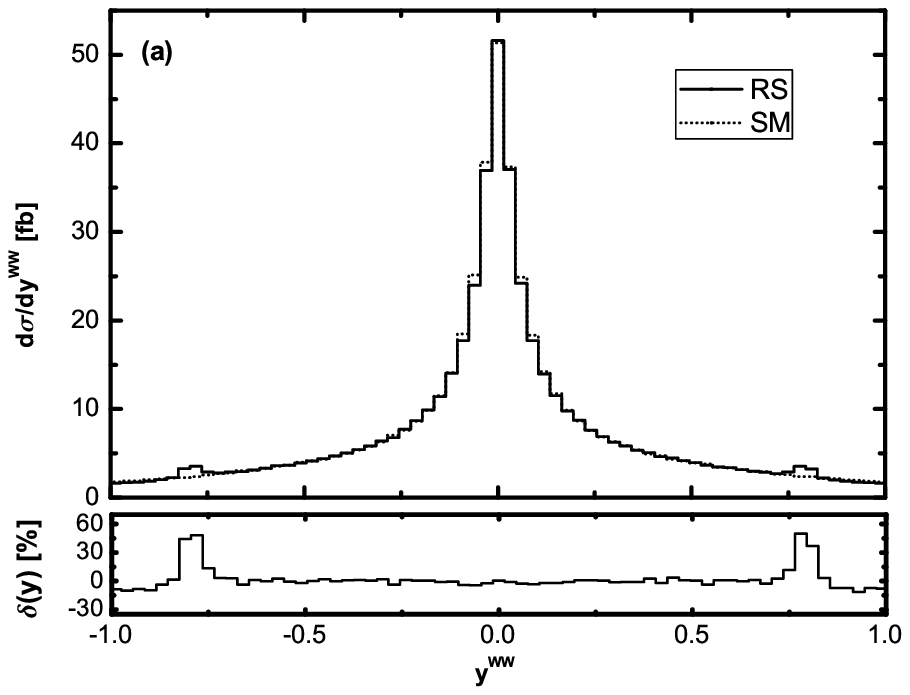}%
\hspace{0in}%
\includegraphics[scale=0.7]{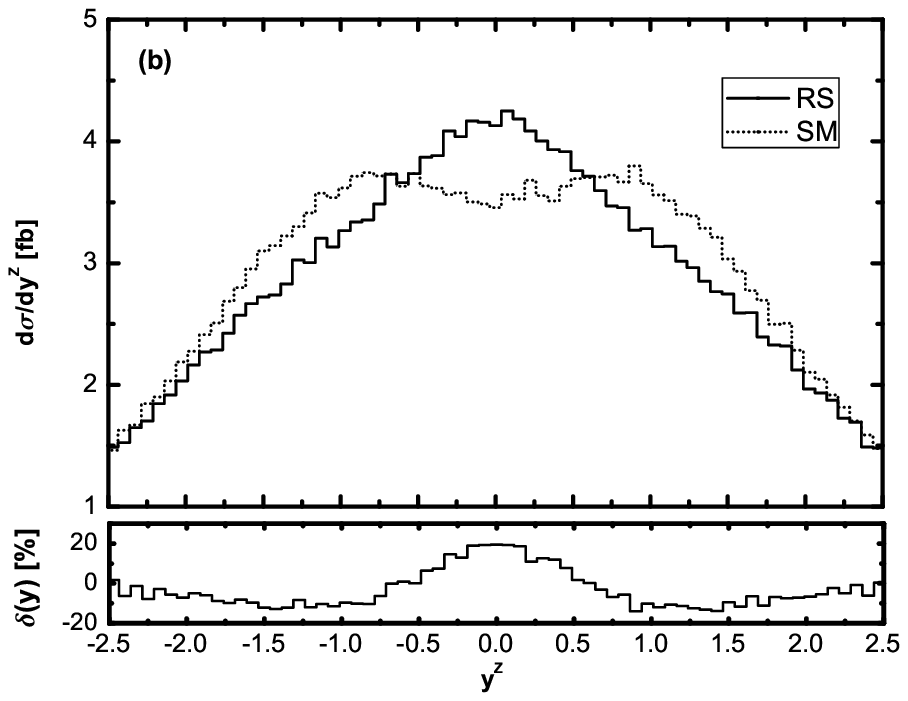}%
\hspace{0in}%
\caption{ \label{fig9} The rapidity distributions of final $W$ pair and
$Z$ and the corresponding relative RS discrepancies for the
$e^+e^- \to W^{+}W^{-}Z$ process in both the SM and the RS model at
the $\sqrt{s}=5$~TeV CLIC, with the RS parameters $M_1=2.25$~TeV and
$c_0=0.1$. (a) for $y^{WW}$, (b) for $y^{Z}$. }
\end{center}
\end{figure}

In Figs.\ref{fig8} and \ref{fig9}, we present the rapidity ($y$)
distributions of the final $W$ pair, $\gamma$ and $Z$ boson for the
$e^+e^- \to W^+W^-\gamma, W^+W^-Z$ processes at the $\sqrt{s}=5$~TeV
CLIC. In each plot of Figs.\ref{fig8}-\ref{fig9}, the $y^{WW}$,
$y^{\gamma}$ and $y^Z$ distributions are given in both the SM and
the RS model for comparison, and the corresponding relative RS
discrepancies, defined as $\delta(y)\equiv
\left(\frac{d\sigma_{RS}}{dy}- \frac{d\sigma_{SM}}{dy}\right)
/\frac{d\sigma_{SM}}{dy}$, are depicted in each nether plot. They
show that the curves of $\delta(y^{WW})$ at the CLIC are quite
different from the corresponding results at the ILC in the LED model
presented in Ref.\cite{14-wwvled} due to the RS KK-graviton
resonance effects. The $\delta(y^{WW})$ distributions at the CLIC
depicted in Figs.\ref{fig8}(a) and \ref{fig9}(a) reach the same
maximum value of about $50\%$ at the positions of $y^{WW}\sim \pm
0.78$. In addition, $\delta(y^{\gamma})$ in Fig.\ref{fig8}(b)
reaches its maximum (minimum) value about $12\%$ ($-12\%$) at
$y^{\gamma}\sim 0$ ($y^{\gamma}\sim \pm 1.0$), and $\delta(y^{Z})$
in Fig.\ref{fig9}(b) achieves its maximum (minimum) value of of
about $19\%$ ($-14\%$) at $y^{Z}\sim 0$ ($y^{Z}\sim \pm 1.4$).

\par
\begin{figure}[htbp]
\begin{center}
\includegraphics[scale=0.7]{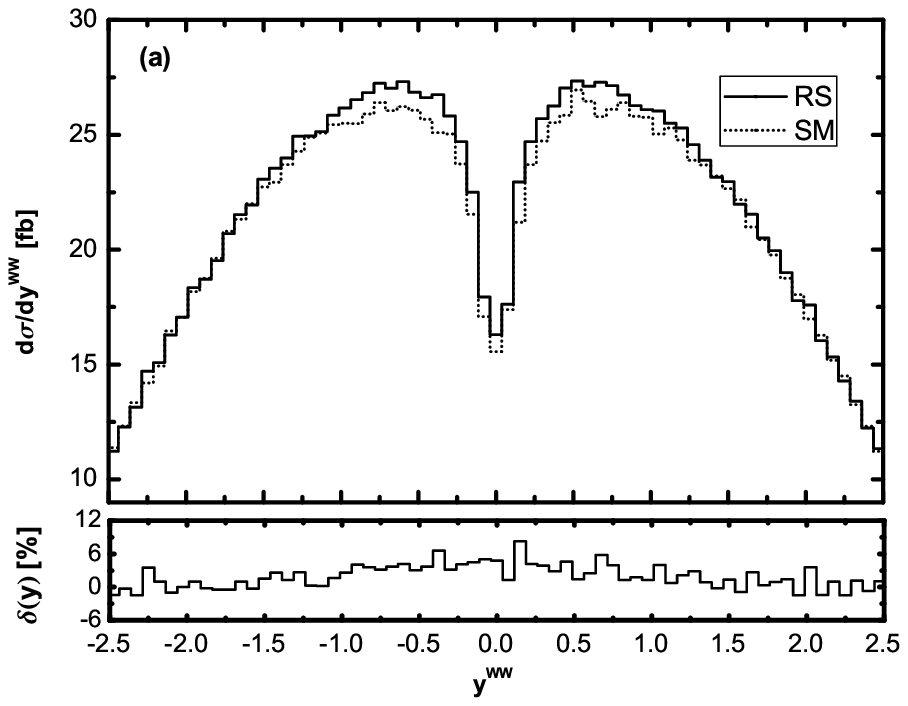}%
\hspace{0in}%
\includegraphics[scale=0.7]{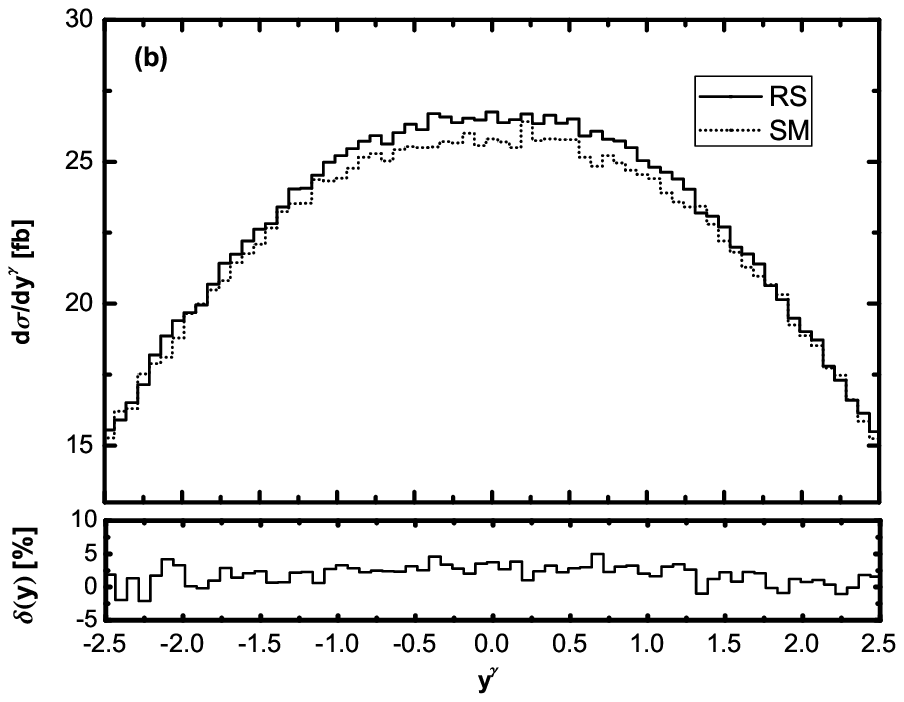}%
\hspace{0in}%
\caption{ \label{fig10} The rapidity distributions of final $W$ pair and
$\gamma$ and the corresponding relative RS discrepancies for the $pp
\to W^{+}W^{-}\gamma$ process in both the SM and the RS model at the
$\sqrt{s}=14$~TeV LHC, with the RS parameters $M_1=2.25$~TeV and
$c_0=0.1$. (a) for $y^{WW}$, (b) for $y^{\gamma}$.  }
\end{center}
\end{figure}
\begin{figure}[htbp]
\begin{center}
\includegraphics[scale=0.7]{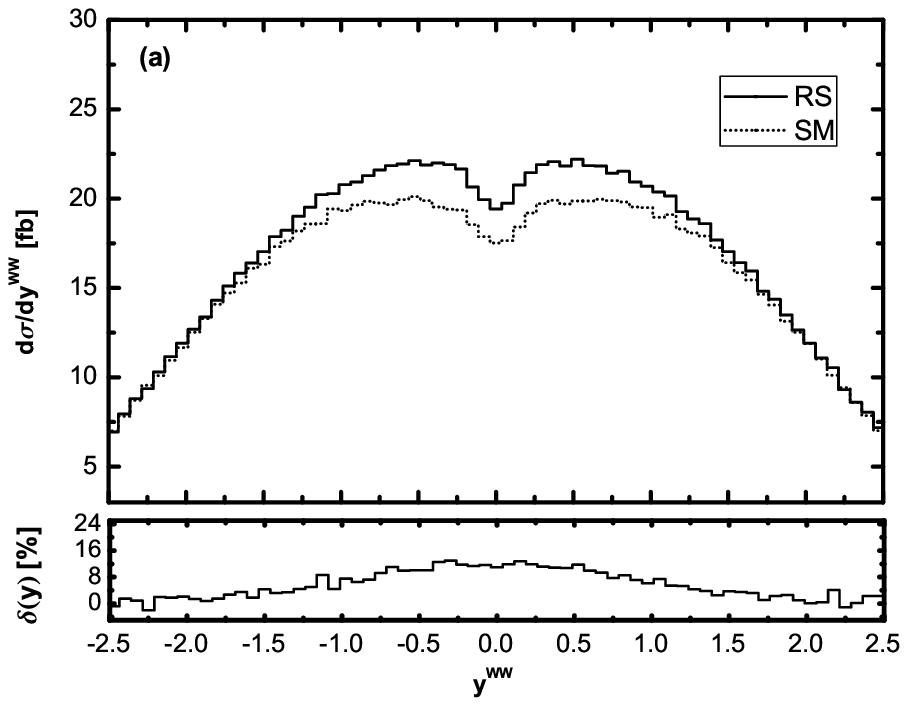}%
\hspace{0in}%
\includegraphics[scale=0.7]{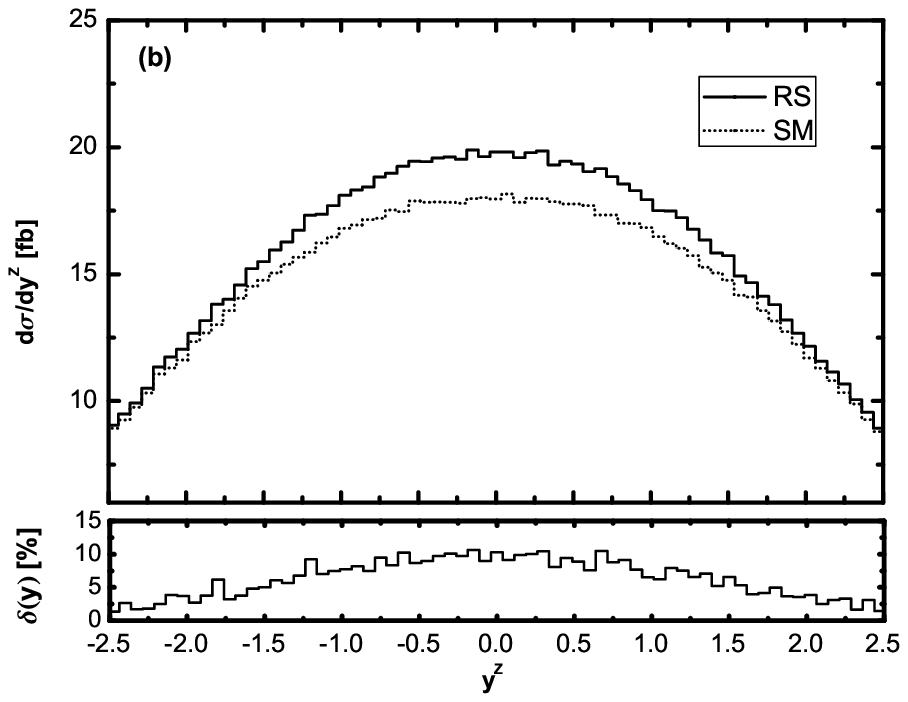}%
\hspace{0in}%
\caption{ \label{fig11} The rapidity distributions of final $W$ pair and
$Z$ and the corresponding relative RS discrepancies for the
$pp \to  W^{+}W^{-}Z$ process in both the SM and the RS model at the
$\sqrt{s}=14$~TeV LHC, with the RS parameters $M_1=2.25$~TeV and
$c_0=0.1$. (a) for $y^{WW}$, (b) for $y^{Z}$.  }
\end{center}
\end{figure}

The $y^{WW}$, $y^{\gamma}$ and $y^Z$ distributions of the $pp \to
W^+W^-\gamma, W^+W^-Z$ processes in the SM and the RS model at the
$\sqrt{s}=14$~TeV LHC are shown in Figs.\ref{fig10} and \ref{fig11},
separately. There we also provide the corresponding relative RS
discrepancies depicted in each nether plot. From the figures we can
see that the line shapes of the relative RS discrepancies at the LHC
differ from those at the CLIC, and the variations of
$\delta(y^{WW})$, $\delta(y^{\gamma})$ and $\delta(y^Z)$ at the LHC
are milder than the corresponding ones at the CLIC shown in
Figs.\ref{fig8} and \ref{fig9}.

\par
\begin{figure}[htbp]
\begin{center}
\includegraphics[scale=0.7]{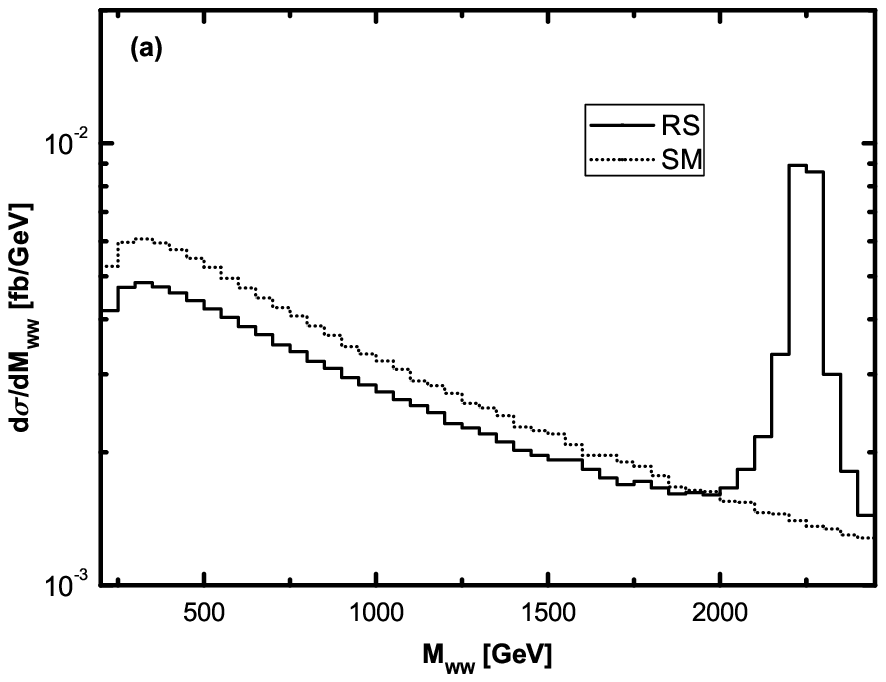}%
\hspace{0in}%
\includegraphics[scale=0.7]{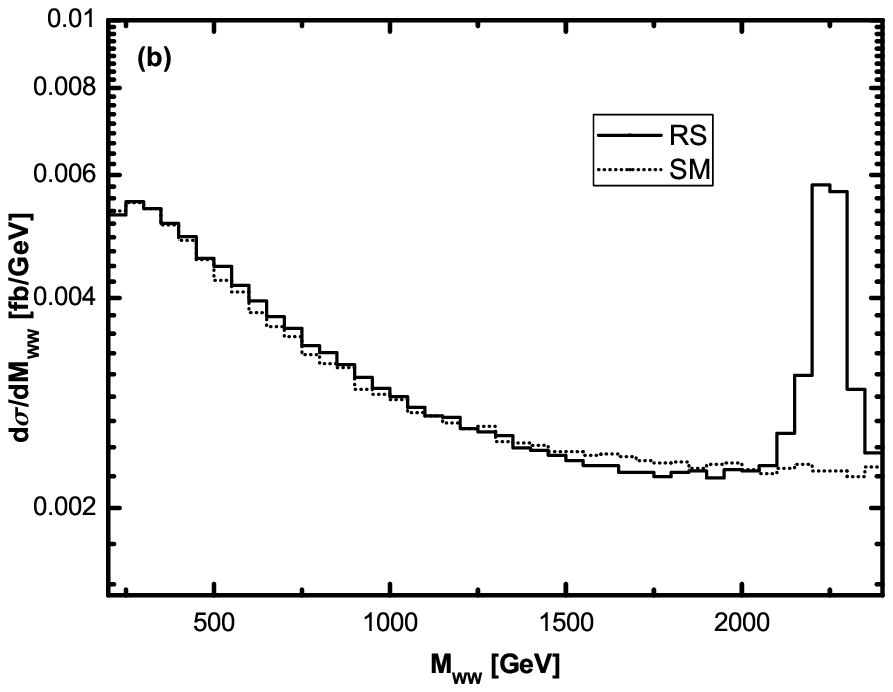}%
\hspace{0in}%
\caption{ \label{fig12}  $M_{WW}$ distributions in both the SM and
the RS model at the $\sqrt{s}=5$~TeV CLIC, with the RS parameters
$M_1=2.25$~TeV and $c_0=0.1$. (a) for the $e^+e^- \to
W^{+}W^{-}\gamma$ process, (b) for the $e^+e^- \to W^{+}W^{-}Z$
process. }
\end{center}
\end{figure}
\begin{figure}[htbp]
\begin{center}
\includegraphics[scale=0.7]{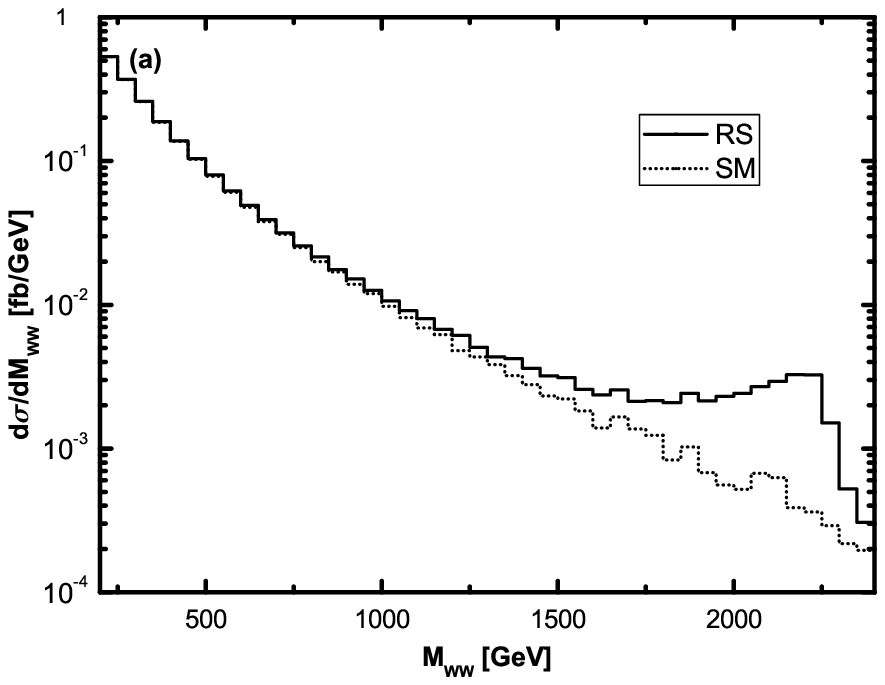}%
\hspace{0in}%
\includegraphics[scale=0.7]{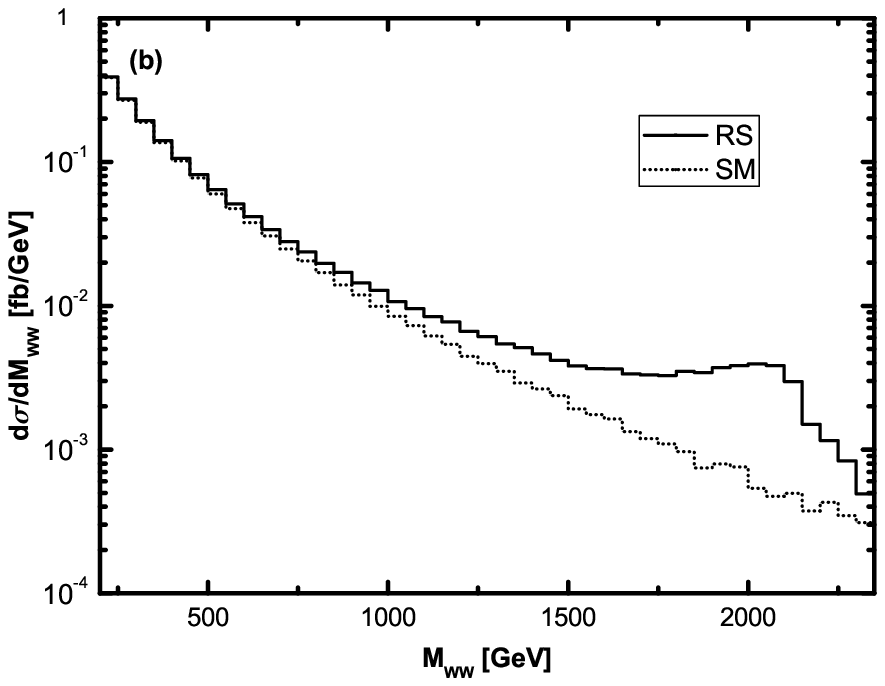}%
\hspace{0in}%
\caption{ \label{fig13}  $M_{WW}$ distributions in both the SM and
the RS model at the $\sqrt{s}=14$~TeV LHC, with the RS parameters
$M_1=2.25$~TeV and $c_0=0.1$. (a) for the $pp \to W^{+}W^{-}\gamma$
process, (b) for the $pp \to W^{+}W^{-}Z$ process. }
\end{center}
\end{figure}
\begin{figure}[htbp]
\begin{center}
\includegraphics[scale=0.7]{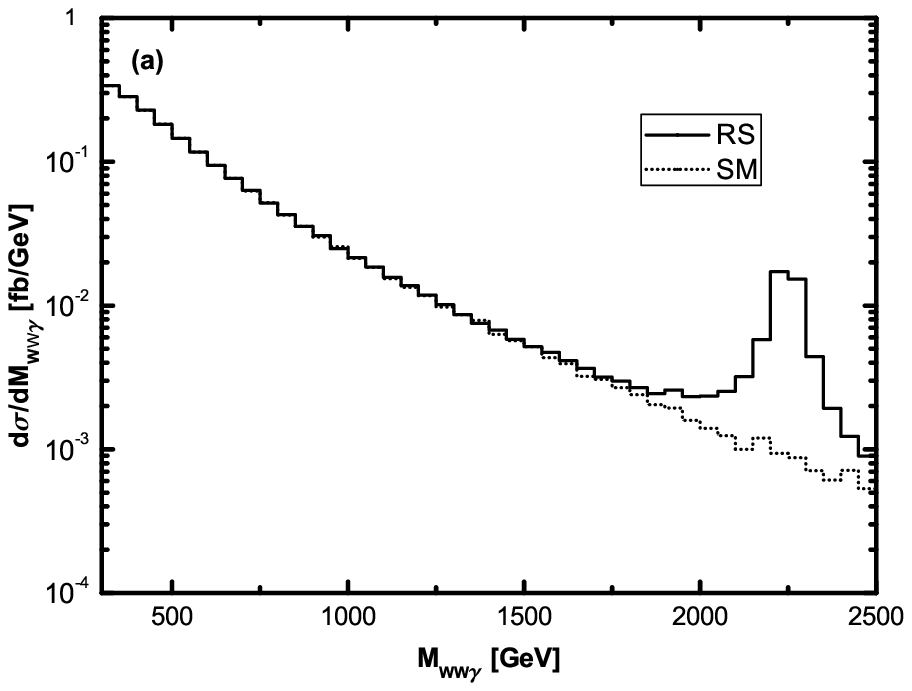}%
\hspace{0in}%
\includegraphics[scale=0.7]{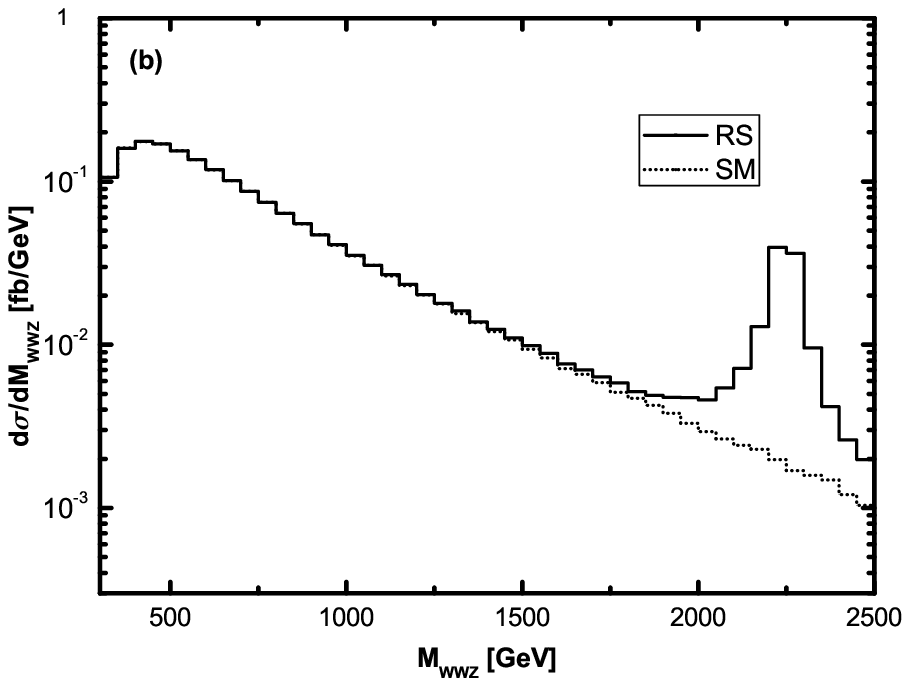}%
\hspace{0in}%
\caption{ \label{fig14}  $M_{WW\gamma(Z)}$ distributions in both
the SM and the RS model at the $\sqrt{s}=14$~TeV LHC, with the RS
parameters $M_1=2.25$~TeV and $c_0=0.1$. (a) for the $pp \to
W^{+}W^{-}\gamma$ process, (b) for the $pp \to W^{+}W^{-}Z$
process. }
\end{center}
\end{figure}

In Figs.\ref{fig12} and \ref{fig13}, we present the $W$-pair
invariant mass ($M_{WW}$) distributions in both the SM and the RS
model at the $\sqrt{s}=5$~TeV CLIC and $\sqrt{s}=14$~TeV LHC,
respectively. In Figs.\ref{fig14}(a) and \ref{fig14}(b), the $W^+W^-\gamma/Z$
invariant mass ($M_{WW\gamma/Z}$) distributions for the $pp \to
W^{+}W^{-}\gamma,W^{+}W^{-}Z$ processes at the LHC are depicted,
separately. In each plot of Figs.\ref{fig12}-\ref{fig14}, the
peaks on the solid curves indicate the existence of the RS
KK graviton. We can see that the RS resonances appear at the
locations of $M_{WW}\simeq M_1=2.25$~TeV in Figs.\ref{fig12}-\ref{fig13} 
and $M_{WW\gamma/Z}\simeq M_1=2.25$~TeV in Fig.\ref{fig14}, where 
$M_1=2.25$~TeV is the mass of the lightest RS KK graviton. 
Figs.\ref{fig12}-\ref{fig14} show that the spin-2 RS KK graviton, 
which couples not only with the $W$ pair but also with the 
$W^+W^-\gamma(Z)$ vertices, contributes dominantly over the 
SM component in the RS KK-graviton resonance region, which is the 
character of the RS model and definitely differs from the results 
in the LED model \cite{14-wwvled}.

\par
\begin{figure}[htbp]
\begin{center}
\includegraphics[scale=0.7]{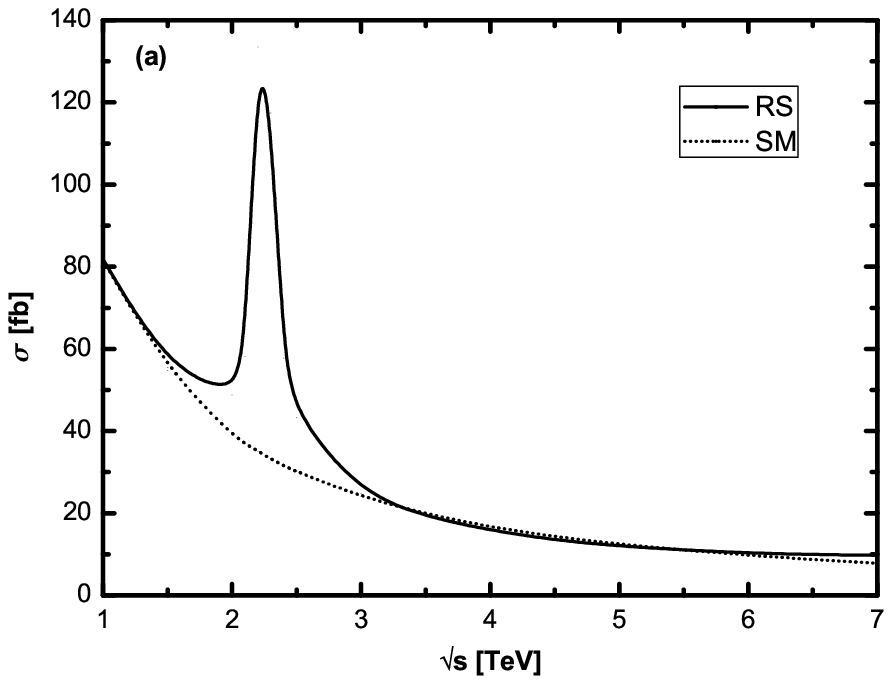}%
\hspace{0in}%
\includegraphics[scale=0.7]{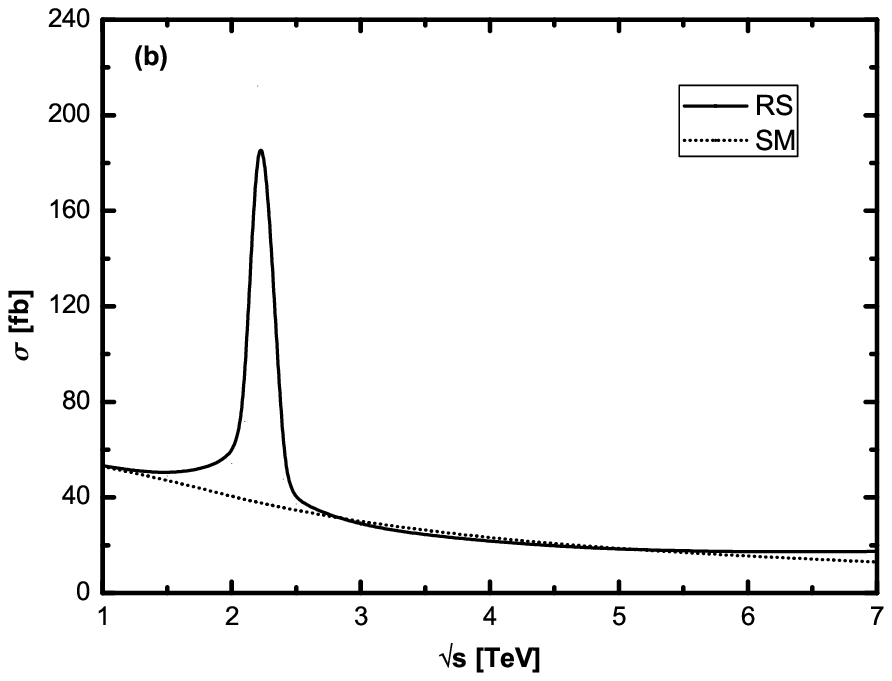}%
\hspace{0in}%
\caption{ \label{fig15} The integrated cross sections as
functions of the c.m.s energy $\sqrt{s}$ in both the SM and the RS
model at the CLIC, with the RS parameters $M_1=2.25$~TeV and $c_0=0.1$.
(a) for the $e^+e^- \to W^{+}W^{-}\gamma$ process,
(b) for the $e^+e^- \to W^{+}W^{-}Z$
process. }
\end{center}
\end{figure}
\begin{figure}[htbp]
\begin{center}
\includegraphics[scale=0.7]{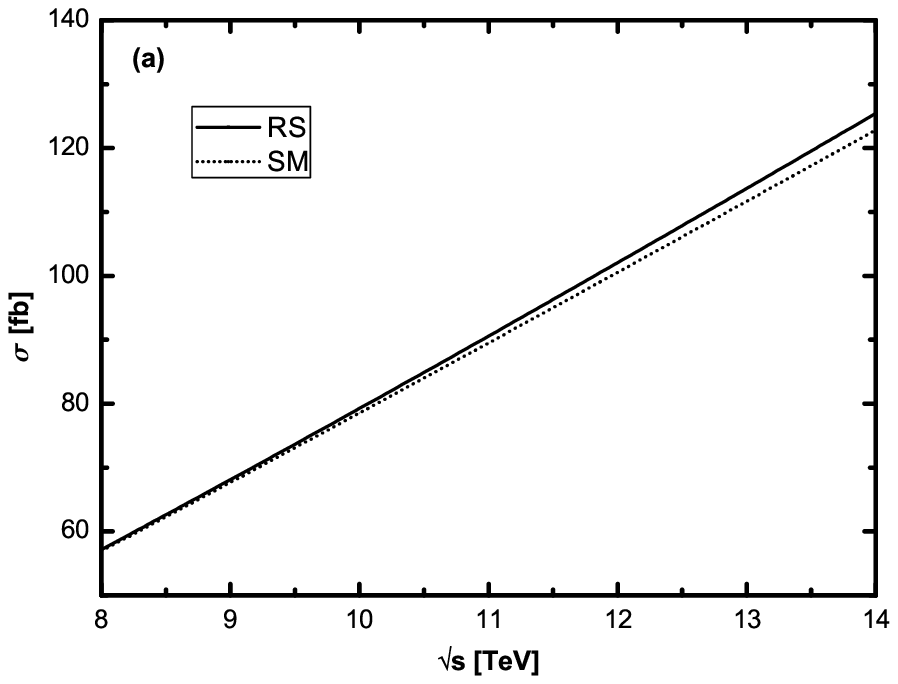}%
\hspace{0in}%
\includegraphics[scale=0.7]{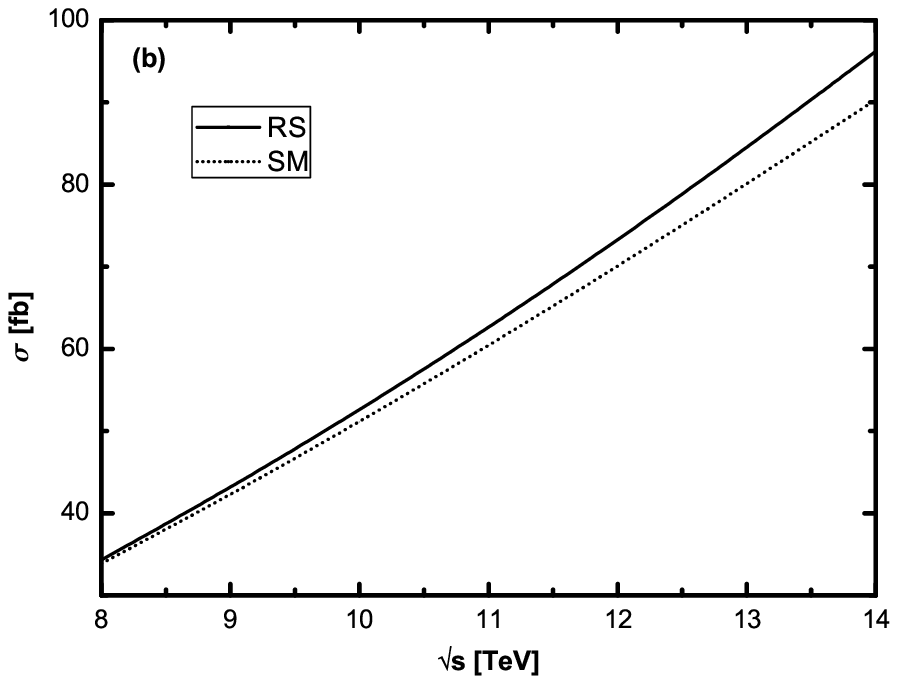}%
\hspace{0in}%
\caption{ \label{fig16} The integrated cross sections as functions of the c.m.s
energy $\sqrt{s}$ in both the SM and the RS model at the LHC, with the RS parameters
$M_1=2.25$~TeV and $c_0=0.1$. (a) for the $pp \to W^{+}W^{-}\gamma$
process, (b) for the $pp \to W^{+}W^{-}Z$ process.}
\end{center}
\end{figure}

In Figs.\ref{fig15} and \ref{fig16}, we present the integrated cross
sections as functions of the c.m.s energy $\sqrt{s}$ at the CLIC and
the LHC, respectively. From Figs.\ref{fig15}(a) and \ref{fig15}(b) we find that the
integrated cross sections at the CLIC in both the SM and the RS
model decrease as $\sqrt{s}$ becomes larger, and there exists an RS
KK-graviton resonance peak on each curve for the $e^+e^- \to
W^{+}W^{-}\gamma, W^{+}W^{-}Z$ processes in the RS model at the
position of $\sqrt{s}\simeq M_1=2.25$~TeV. By contrast, the curves
in Fig.\ref{fig16} for the $pp \to W^{+}W^{-}\gamma, W^{+}W^{-}Z$
processes at the LHC in both the SM and RS model increase with the
increment of $\sqrt{s}$, and there is no appearance of the RS
resonance peak due to the convolution of the PDFs.

\par
\begin{figure}[htbp]
\begin{center}
\includegraphics[scale=0.7]{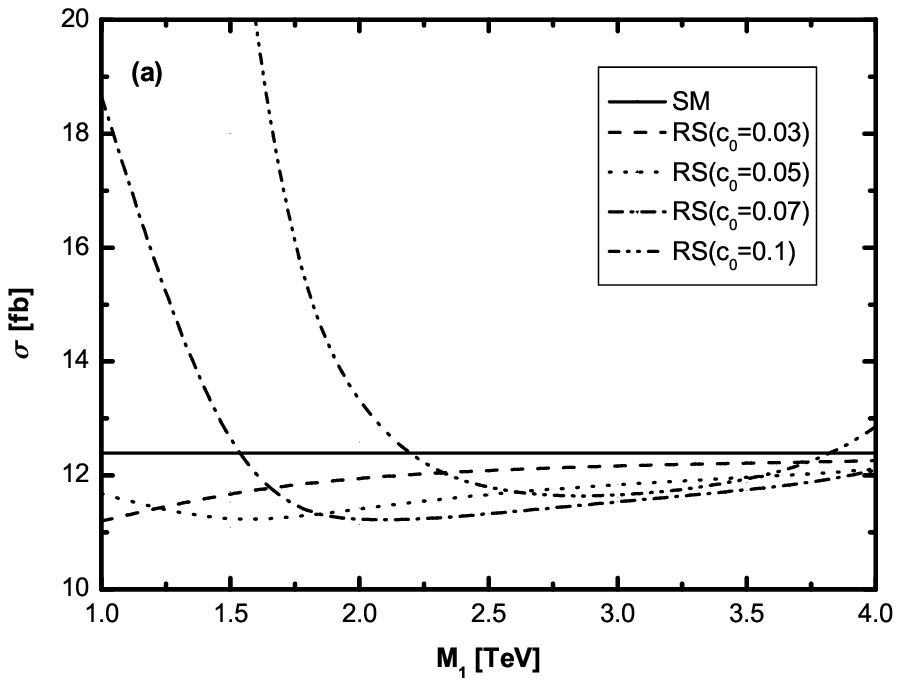}%
\hspace{0in}%
\includegraphics[scale=0.7]{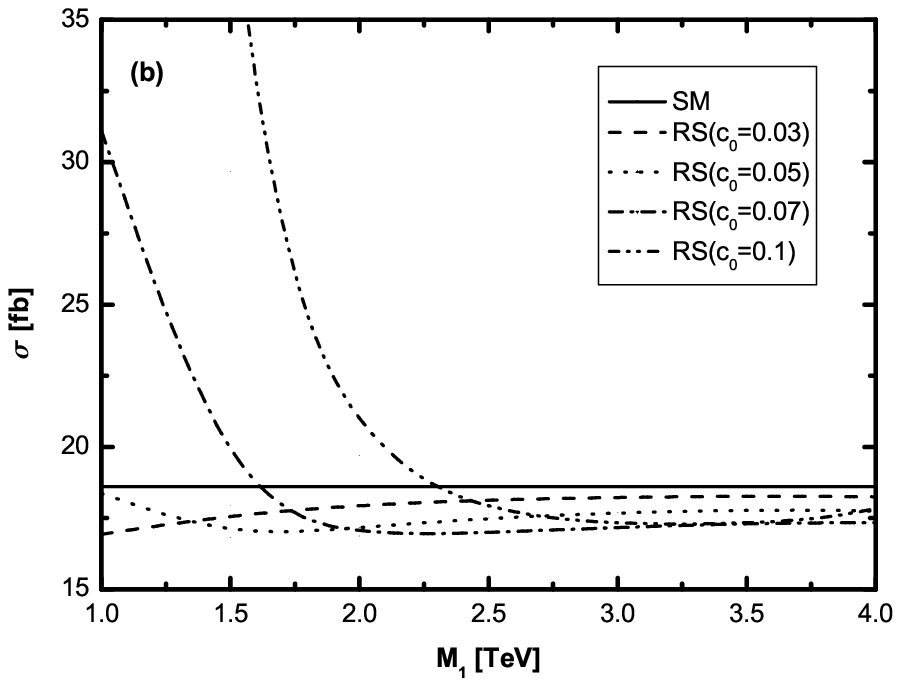}%
\hspace{0in}%
\caption{ \label{fig17} The integrated cross sections as functions of
$M_1$ with $c_0=0.03,0.05,0.07$ and $0.1$ at the $\sqrt{s}=5$~TeV
CLIC. The SM results appear as the straight lines. (a) for the
$e^+e^- \to W^{+}W^{-}\gamma$ process, (b) for the $e^+e^- \to
W^{+}W^{-}Z$ process. }
\end{center}
\end{figure}
\begin{figure}[htbp]
\begin{center}
\includegraphics[scale=0.7]{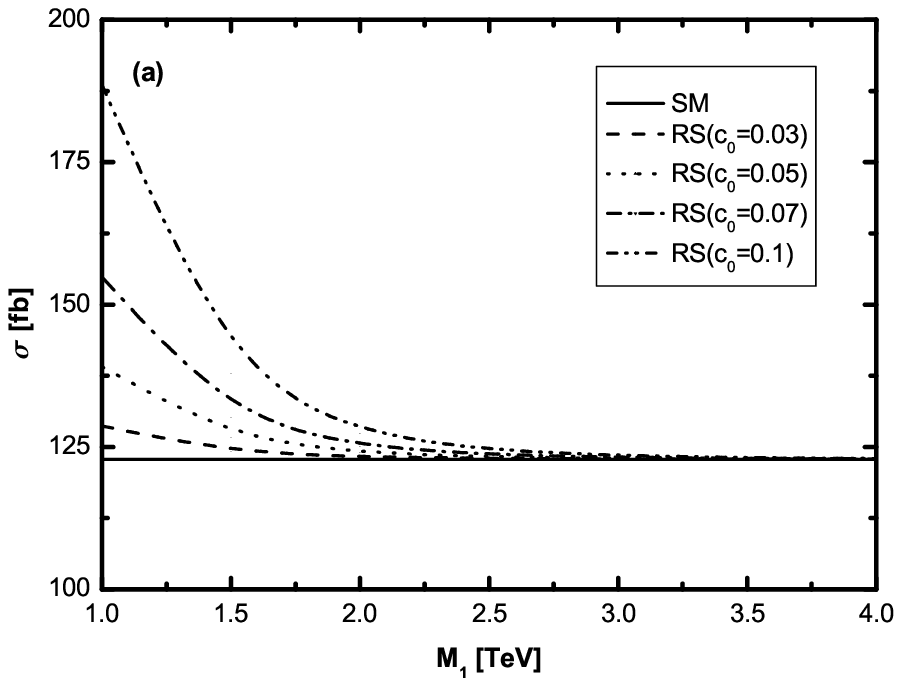}%
\hspace{0in}%
\includegraphics[scale=0.7]{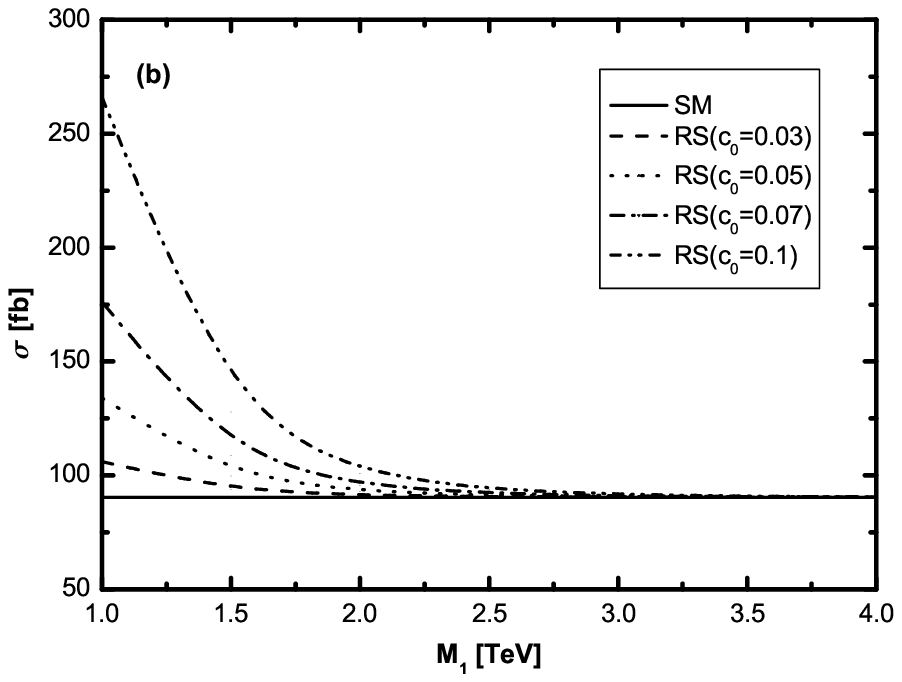}%
\hspace{0in}%
\caption{ \label{fig18} The integrated cross sections as functions of
$M_1$ with $c_0=0.03,0.05,0.07$ and $0.1$ at the $\sqrt{s}=14$~TeV LHC.
The SM results appear as the straight lines. (a) for the $pp \to
W^{+}W^{-}\gamma$ process, (b) for the $pp \to W^{+}W^{-}Z$ process.
}
\end{center}
\end{figure}

In Figs.\ref{fig17} and \ref{fig18}, we show the relations between
the integrated cross sections and the RS parameter $M_1$ with
$c_0=0.03,0.05,0.07$ and $0.1$, in the $W^{+}W^{-}\gamma(Z)$ production 
processes at the CLIC and LHC, respectively.
The horizon line in each plot stands for the SM cross section,
which is independent of $M_1$ and $c_0$. Compared with the obscure
behaviors of $\sigma(M_1,c_0)$ at the CLIC shown in
Fig.\ref{fig17}, the cross sections for the $pp \to
W^{+}W^{-}\gamma, W^{+}W^{-}Z$ processes in Fig.\ref{fig18}
exhibit more monotone relationship with the increment of the RS
parameter $M_1$ or $c_0$. Figure.\ref{fig18} shows when the value of
$M_1$ is fixed, we see the larger the value of $c_0$, the more
evident the deviation due to the large RS KK-graviton contributions. In addition, 
it shows $\sigma(M_1,c_0)$ decreases and gradually approaches the SM
results with the increment of $M_1$.

\par
From the above discussion we can see that the kinematical
observables for the $W^+W^-\gamma(Z)$ production processes at the
CLIC and LHC have generally different behaviors, and the resonance
effects of the RS KK graviton at both colliders are also
dissimilar to those in the LED model shown in
Ref.\cite{14-wwvled}. This difference can be ascribed to the
following distinct features of the RS KK-graviton spectrum
\cite{22-RSbible}: (1) In the LED model the multiplicity of the
LED KK-graviton density collectively contributes to arrive at the
electroweak scale, while the coupling of the RS KK graviton with
the SM particles individually reaches the electroweak strength via
the enhancement of the warp factor $e^{\pi {\cal K} R_c}$. (2) The
LED KK-graviton spectrum is closely spaced with mass separation 
$\Delta m \sim {\cal O}(1/R) <0.1$~GeV leading to
form a quasicontinuum, which manifests itself as a nonresonance
contribution. Comparatively, the spectrum of the RS KK graviton is
widely spaced with mass splitting of TeV order, and can possibly make
the RS KK graviton being produced as resonance at the LHC
and the CLIC.

\vskip 5mm
\section{SUMMARY}
\par
In this paper, we study the effects of the virtual RS KK graviton
on the $W^{+}W^{-}\gamma$ and $W^{+}W^{-}Z$ productions at the LHC
and the CLIC. The SM background is also included for comparison.
We find that the transverse momentum ($p_{T}$) distributions and
the corresponding RS relative discrepancies $\delta(p_T)$ at the
CLIC and the LHC exhibit opposite behaviors, namely, the RS
KK-graviton effects tend to eliminate the SM contributions at the
CLIC, but enhance them at the LHC. We provide the rapidity
distributions ($y$) of the final particles, and it is shown that 
the RS relative discrepancies $\delta(y)$
at the LHC differ from those at the CLIC, and the $\delta(y)$
distribution shapes are unlike those in the LED model
\cite{14-wwvled}. The invariant mass ($M_{WW}$ or
$M_{WW\gamma/Z}$) distributions in the RS model are also
presented. There exists a resonance peak on each distribution, and
the RS KK-graviton makes dominant contributions over the SM
background in the KK graviton resonant region. Moreover, we study
the effects of the colliding energy $\sqrt{s}$ and the RS
parameters on the integrated cross sections. We find that the
results for the $W^+W^-\gamma(Z)$ production processes in the RS
model exhibit distinct behaviors from those in the LED
model due to the nonfactorizable coupling property and the 
sufficiently separated resonance characteristic of the RS KK-graviton 
spectrum. We conclude that the CLIC with unprecedented precision
and high colliding energy has a potential advantage over the LHC in
studying the phenomenological effects of the RS KK graviton on the
$W^{+}W^{-}\gamma(Z)$ productions.

\vskip 5mm
\par
\noindent{\large\bf Acknowledgments:} This work was supported in
part by the National Natural Science Foundation of China
(No.11075150, No.11005101, No.11275190) and the Fundamental Research
Funds for the Central Universities (No.WK2030040024).

\par
{\bf Appendix: The relevant couplings }
\par
The Feynman rules for the vertices in the RS model
related to our calculations are listed below
\cite{7-ppGkk, 8-ppllr, 22-RSbible}.
\begin{itemize}
\item[(i)]
$G_{\rm KK}^{\mu
\nu}(k_3)-\bar{\psi}(k_1)-\psi(k_2)~\textrm{vertex}: $
\begin{eqnarray}
-i {1 \over {4 \Lambda_\pi}} \left[\gamma^{\mu} (k_1 + k_2)^{\nu} +
\gamma^{\nu} (k_1 + k_2)^{\mu} - 2 \eta^{\mu \nu} (\rlap/{k}_1 +
\rlap/{k}_2 - 2 m_{\psi}) \right]
\end{eqnarray}
\item[(ii)]
$G_{\rm KK}^{\mu
\nu}(k_4)-\bar{\psi}(k_1)-\psi(k_2)-A^{\rho}(k_3)~\textrm{vertex}:
$
\begin{eqnarray}
i e Q_{f} {1 \over {2 \Lambda_\pi}} \left( \gamma^{\mu} \eta^{\nu
\rho} + \gamma^{\nu} \eta^{\mu \rho} - 2 \gamma^{\rho}\eta^{\mu \nu}
\right)
\end{eqnarray}
\item[(iii)]
$G_{\rm KK}^{\mu
\nu}(k_4)-\bar{\psi}(k_1)-\psi(k_2)-Z^{\rho}(k_3)~\textrm{vertex}:
$
\begin{eqnarray}
-i e {1 \over {2 \Lambda_\pi}} \left[(\gamma^{\mu} \eta^{\nu
\rho} + \gamma^{\nu} \eta^{\mu \rho} - 2 \gamma^{\rho}\eta^{\mu \nu})
(\upsilon_f -a_f \gamma_5) \right]
\end{eqnarray}
\item[(iv)]
$G_{\rm KK}^{\mu \nu}(k_3)-W^{+ \rho}(k_1)-W^{-
\sigma}(k_2)~\textrm{vertex}: $
\begin{eqnarray}
-2 i {1 \over \Lambda_\pi} \left[B^{\mu \nu \rho \sigma} m_W^2 +
(C^{\mu \nu \rho \sigma \tau \beta} - C^{\mu \nu \rho \beta \sigma
\tau}) k_{1\tau} k_{2\beta} + \frac{1}{\xi}E^{\mu \nu \rho
\sigma}(k_1,k_2)\right]
\end{eqnarray}
\item[(v)]
$G_{\rm KK}^{\mu \nu}(k_4)-W^{+ \rho}(k_1)-W^{-\sigma}(k_2)
-A^{\lambda}(k_3)~\textrm{vertex}: $
\begin{eqnarray}
-2 i e {1 \over \Lambda_\pi} \left[(k_1-k_3)_{\tau}C^{\mu
\nu \tau \sigma \rho \lambda}+ (k_2-k_1)_{\tau}C^{\mu \nu \sigma
\rho \tau \lambda} + (k_3-k_2)_{\tau}C^{\mu \nu \lambda \sigma \tau
\rho}\right]
\end{eqnarray}
\item[(vi)]
$G_{\rm KK}^{\mu \nu}(k_4)-W^{+ \rho}(k_1)-W^{-\sigma}(k_2)
-Z^{\lambda}(k_3)~\textrm{vertex}: $
\begin{eqnarray}
2 i e \frac{s_w}{c_w} {1 \over \Lambda_\pi} \left[(k_1-k_3)_{\tau}C^{\mu
\nu \tau \sigma \rho \lambda}+ (k_2-k_1)_{\tau}C^{\mu \nu \sigma
\rho \tau \lambda} + (k_3-k_2)_{\tau}C^{\mu \nu \lambda \sigma \tau
\rho}\right]
\end{eqnarray}
\end{itemize}
where $G_{\rm KK}^{\mu \nu}$, $\psi$, $W^{\pm \mu}$, $Z^{\mu}$ and
$A^{\mu}$ refer to the fields of the RS KK graviton, fermion,
$W$ boson, $Z$ boson and photon, respectively. We assume all the
momenta flow into the vertices except for the fermion momenta, which are
set along with the fermion line directions. The electric
coupling strength $e=\sqrt{4\pi\alpha}$, $\alpha$ is the
fine-structure constant, $Q_f$ is the electric charge, $s_w~(c_w)$
are sine (cosine) of the Weinberg angle, the vector and axial vector
couplings of the $Z$ boson, i.e., $\upsilon_f$ and $a_f$, are the
same as those in the SM. We adopt the Feynman gauge and the gauge-fixing
parameter is then set as $\xi=1$. The tensor coefficients $B^{\mu
\nu \alpha \beta}$, $C^{\rho \sigma \mu \mu \alpha \beta}$ and
$E^{\mu \nu \rho \sigma}(k_{1},k_{2})$ are defined as
\cite{14-wwvled}
\begin{eqnarray}
B^{\mu \nu \alpha \beta} & = & \frac{1}{2}
      (\eta^{\mu \nu}\eta^{\alpha \beta}
      -\eta^{\mu \alpha}\eta^{\nu \beta}
      -\eta^{\mu \beta}\eta^{\nu \alpha}),
       \nb \\
C^{\rho \sigma \mu \nu \alpha \beta} & = & \frac{1}{2}
      [\eta^{\rho \sigma}\eta^{\mu \nu}\eta^{\alpha \beta}
     -(\eta^{\rho \mu}\eta^{\sigma \nu}\eta^{\alpha \beta}
      +\eta^{\rho \nu}\eta^{\sigma \mu}\eta^{\alpha \beta}
      +\eta^{\rho \alpha}\eta^{\sigma \beta}\eta^{\mu \nu}
      +\eta^{\rho \beta}\eta^{\sigma \alpha}\eta^{\mu \nu})],
      \nb \\
E^{\mu \nu \rho \sigma}(k_{1},k_{2}) & = &
      \eta^{\mu \nu}(k_1^{\rho} k_1^{\sigma} + k_2^{\rho} k_2^{\sigma}
      + k_1^{\rho} k_2^{\sigma}) - \left [\eta^{\nu \sigma} k_1^{\mu} k_1^{\rho}
      + \eta^{\nu \rho} k_2^{\mu} k_2^{\sigma} + (\mu \leftrightarrow \nu)\right ]. \nb
\end{eqnarray}

\end{document}